\documentclass[a4paper,11pt]{article}
\usepackage{jheppub} 
\usepackage{multirow}
\usepackage{makecell}
\usepackage{fontawesome5}

\title{\boldmath Notes on the bootstrap of four-point conformal integrals}

\author[a]{Xuhang Jiang}\emailAdd{xhjiang@itp.ac.cn}
\affiliation[a]{Institute of Theoretical Physics, Chinese Academy of Sciences, Beijing 100190, China}

\abstract{
    We set up a bootstrap workflow to study four-point conformal integrals in position
space, using leading singularities, single-valued multiple polylogarithmic
ansätze and boundary data from expansion by regions. These four-point conformal integrals are general in the sense that they are generated by the four-point projections of all possible $f$-graphs, including
all non-planar $f$-graph sectors. For three-loop cases, fourteen of the fifteen inequivalent integrand basis can be directly calculated by \texttt{HyperlogProcedures} and the last one is fixed by Gram identity. Then we concentrate on how far the bootstrap workflow can go for four-loop cases, though it works for three-loop cases as well.
We show that integrals with
several leading singularities can be made tractable by decomposing them into
pieces with simpler cut structure. Some four-loop integrals which can not be calculated or very hard to be calculated by other methods for now are obtained in this way. We also provide a package with skill files which is suitable to be read and used by current AI models.}

\begin{document}
\maketitle
\flushbottom

\section{Introduction}
\label{sec:intro}
Conformal integrals in position space form a distinguished class of
Feynman integrals appearing in correlation functions of conformal field
theories, most notably in perturbative $\mathcal N=4$ super-Yang--Mills theory. Under the inversion map, we have
\begin{equation}
    x_{i}^{\mu}\to \frac{x_{i}^{\mu}}{x_{i}^2}, \quad (x_i-x_j)^2\to \left(\frac{x_i}{x_i^2}-\frac{x_j}{x_j^2}\right)^2=\frac{x_{ij}^2}{x_i^2x_j^2}
\end{equation}
Thus their conformal invariance implies that the degree of each internal vertex (which is to be integrated out) appearing in the integrand is 4\footnote{For example, $1/x_{15}^2$ has degree 1 for vertex $5$ and $\frac{x_{15}^2}{x_{25}^2x_{35}^2}$ has degree $1=2-1$ for vertex $5$ as well.}.
In this note, we concentrate on four-point conformal integrals, which naturally contribute to the four-point correlation functions~\cite{Eden:2011we,Heslop:2022xgp}. Due to the same reason of conformal invariance, after fixing the four external points to be $x_1,x_2,x_3,x_4$, the integrated results only depend on the two cross ratios
\begin{equation}
    u\equiv \frac{x_{12}^2x_{34}^2}{x_{13}^2x_{24}^2}=z\bar{z}, \quad v\equiv \frac{x_{14}^2x_{23}^2}{x_{13}^2x_{24}^2}=(1-z)(1-\bar{z}).
\end{equation}
$z,\bar{z}$ are the complex parameterization of these two cross ratios. For four points in the complex plane we can always fix three points to $0,1,\infty$, the remaining complex variable can be taken to be $z$. Then $u$ and $v$ are two nontrivial distances between $0,1,z$. Under the permutation of external vertices, $u,v$ may have different relations with $z,\bar{z}$, they differ by a M\"obius transformation of $z,\bar{z}$. Four-point conformal integrals are also studied as graphical functions~\cite{Schnetz:2013hqa,Borinsky:2021gkd,Schnetz:2021ebf,Schnetz:2025opm,Chakraborty:2026ziu} in the literature, due to their graphical representations. We can depict each denominator $1/x_{ij}^2$ as a solid edge and each numerator $x_{ij}^2$ as a dashed edge as long as the integrand is given as single rational functions of $x_{ij}^2$.  Especially, many four-point conformal integrals with or without numerators in even dimensions with $d\ge 4$ can be efficiently calculated with the graphical function method, though they are developed initially for the periods in $\phi^4$ theory~\cite{Broadhurst:1995km,Schnetz:2008mp}. This method has been implemented into the maple package \texttt{HyperlogProcedures}~\cite{hyperlog}.

On the one hand, we have several powerful methods to calculate general Feynman integrals, for example, the (canonical) differential equation method for general Feynman integrals~\cite{Kotikov:1990kg,Henn:2013pwa,Caron-Huot:2014lda}, or the iterated integration technique for linear reducible integrals~\cite{Brown:2008um,Panzer:2014caa,Kardos:2026eun} and Mellin-Barnes parameterization method~\cite{Smirnov:2012book}. However, they will either depend heavily on Integration-By-Parts (IBP) identities~\cite{Chetyrkin:1981qh,Laporta:2000dsw,Laporta:2003jz} or the explicit parameterizations of the integrands, both of which become very complicated for higher-loop integrals since they rely heavily on integrand information. On the other hand, the bootstrap method exploited in~\cite{Drummond:2013nda} for three-loop integrals depends only on limited information of the integrands, which can be tackled without requiring large computational resources. It directly tackles the final results which may be very simple. This is why we want to explore further how far the bootstrap method can go for four-loop conformal integrals. We are also interested in how to extract analytic information from the integrands as much as possible.

The first main result of this paper is to show that such a bootstrap workflow indeed works for some conformal integrals which can not be calculated or very hard to be calculated by other methods. The input we use is much weaker than a direct integration. We extract rational leading-singularity prefactors from cuts of the position-space integrand, construct ansätze in single-valued function spaces following these prefactors, and fix the free parameters from boundary data obtained by expansion by regions.
This strategy is relatively effective for rigid polylogarithmic integrals, where the leading singularities already give strong information about the possible function space. The decomposition of integrals into single leading singularity pieces turns out to be useful in this strategy. And finally, we indeed get some new results for four-loop conformal integrals.

The second result, which is a byproduct of this work, is the classification of general conformal integrals which may be generated from $f$-graphs at three and four loops. In short, they are all possible conformal integrals which may (though not necessarily) contribute to $\mathcal{N}=4$ SYM. 
We systematically exploit additional constraints that come from permutation symmetries, Gram-determinant relations, magic identities to classify them first.
At three loops there are four $f$-graphs, one planar and three non-planar. Their four-point projections give fifteen inequivalent conformal integrals. It is interesting that we can determine all of them analytically.
Fourteen of the fifteen basis can be directly calculated by \texttt{HyperlogProcedures} and the last one integral is fixed through a Gram-determinant identity. After including two magic identities and the relevant permutation symmetries, the complete three-loop sector is organized in terms of twelve independent functions.
The three-loop results also provide a controlled testing ground for the
bootstrap method. They allow us to compare the leading-singularity ansatz with
known analytic data, to test how generalized single-valued letters enter the
answer, and to determine how much boundary information is needed to fix the
coefficients. 
At four loops, the power of additional constraints from permutation symmetries, magic identities and Gram identities turns out to be much weaker as the integral basis grows rapidly. 
Instead, leading-singularity analysis gives a systematic way to further classify
integrals and to identify the sectors where a polylogarithmic bootstrap is
plausible. Here the space of conformal integrals is much
larger and more varied. Many integrals have several leading singularities,
non-uniform weight components, or cuts suggesting elliptic behavior, so one
can not expect a universal polylogarithmic bootstrap for the entire four-loop
sector.

This note is organized as follows. In Sec.~\ref{sec:analytic}, we will set up the full bootstrap workflow from the very beginning. In Sec.~\ref{sec:ls}, we discuss how to extract the leading singularities of conformal integrals and exploit this information to construct integrals with single leading singularity. We also discuss how to probe the function space involved in the final results. In Sec.~\ref{sec:boundary}, the calculation of boundary conditions is set up. Since we do not restrict ourselves to planar integrals, we include the four nonplanar two-point master integrals appearing in the boundary calculation. In Sec.~\ref{sec:series}, we discuss how to match the ansatz with boundary conditions, there are still subtleties in this step. Finally, we set up our playground in Sec.~\ref{sec:conformal}. We discuss the conformal integrals to be tested with the bootstrap workflow and their relations to physical quantities. We obtain some new results for four-loop conformal integrals which can not be (or very hard to be) calculated by other methods for now.
\section{Analyticity of Conformal integrals}\label{sec:analytic}

In this section, we focus on the analytic properties of conformal integrals, including their singularities, their boundary values (behavior in singular kinematic limits). We will see that these structures are strong enough to fix the analytic form of these conformal integrals. Though most of the methods and ideas have appeared in the literature long ago, they are sharpened here in the following several aspects: First (mainly discussed in Sec.~\ref{sec:ls}), we have not only further developed the leading singularity analysis for conformal integrals, but also applied ideas of construction of $\mathrm{d}\log$ integrals~\cite{Chicherin:2018old,Herrmann:2019upk,Henn:2020lye,Chen:2020uyk} for some of them. 
Such ideas have been greatly explored in the study of chambers of correlahedron~\cite{He:2025rza} and construction of conformal UT basis for two-loop five-point correlators as well~\cite{Kuo:2025bdr}. However, here this construction is designed for a single conformal integral with a graphical expression (or a graphical function). Furthermore, we also bring up an empirical rule for the ansatz of functions which follows the leading singularity. Second, (mainly discussed in Sec.~\ref{sec:series}), we have further explored how the analytic property of the final results can put much stronger constraints than it appears to. We have obtained the expressions for some four-loop conformal integrals which can not be (or very hard to be) calculated by other methods. At last, we find some tedious technical details are more easily handled by large language model (LLM) than human, thus we also provide a package\footnote{The package is at github \faGithub\, \url{https://github.com/windfolgen/svbwalkthrough.git}.} along with skill files which can be easily read and used by current AI models\footnote{The skill is developed first in the AI research platform \texttt{aether}~\cite{aether} with large language model \texttt{deepseek-v4-pro}~\cite{deepseek}. Different parts are refined with \texttt{gemini-3.5-flash}~\cite{gemini} and \texttt{GLM-5.2}~\cite{glm52}. The integral results calculated with the assistant of large language model are cross-checked by human.}.  
\subsection{Leading Singularity Analysis for Conformal Integrals}\label{sec:ls}
In this section, we give our procedures to analyze the leading singularities of conformal integrals in vertex coordinate space. The basic ideas are the same as in \cite{Drummond:2013nda}. We will further develop this leading singularity method for finite conformal integrals in this section.
\subsubsection{Short review for leading singularity analysis}
\begin{figure}[htbp]
    \centering
    \includegraphics[width=0.3\textwidth]{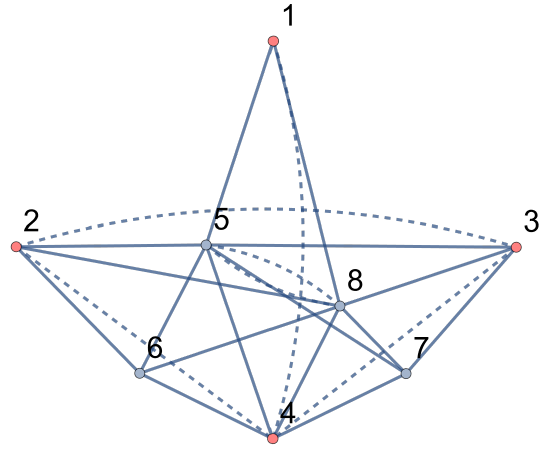}
    \caption{One four-loop conformal integral used to show the standard way to analyze its leading singularity}\label{fig:I4L168}
\end{figure}

We first review with a standard example, a four-loop conformal integral which contains elliptic cuts and will contribute to four-point correlators in $\mathcal{N}=4$ super-Yang-Mills theory~\cite{He:2025rza}. This example shows several essential features we will meet for four-loop conformal integrals.
Its integrand takes the form
\begin{equation}
    I_{168}^{(4)}=\frac{x_{14}^2 x_{23}^2 x_{24}^2 x_{34}^2 x_{58}^4}{x_{15}^2 x_{18}^2 x_{25}^2 x_{26}^2 x_{28}^2 x_{35}^2 x_{37}^2 x_{38}^2 x_{45}^2 x_{46}^2 x_{47}^2 x_{48}^2 x_{56}^2 x_{57}^2
   x_{68}^2 x_{78}^2} .
\end{equation}
and it is depicted in Fig.~\ref{fig:I4L168}. Here we use $I^{(L)}_{i}$ to denote the integrand of an $L$-loop integral with $i$ its index. The corresponding integral is denoted by $\mathcal{I}^{(L)}_{i}$. Sometimes we will suppress the superindex $^{(L)}$ for simplicity if there is no ambiguity.
We can first cut $x_{6}$ and $x_{7}$ out since they are attached to only four points and not connected with each other. They involve the following 8 propagators:
\begin{equation}\label{eq:cut67}
    {\text{cut: }} x_{26}^{2}, \, x_{46}^{2}, \, x_{56}^{2}, \, x_{68}^{2}, \, x_{37}^{2}, \, x_{47}^{2}, \, x_{57}^{2}, \, x_{78}^{2} .
\end{equation}
Then we arrive at
\begin{equation}\label{eq:cut168int}
    I^{\prime}_{168}=\frac{x_{14}^{2}x_{23}^{2}x_{24}^{2}x_{34}^{2}x_{58}^{4}}{x_{15}^2 x_{18}^2 x_{25}^2 x_{28}^2 x_{35}^2 x_{38}^2 x_{45}^2 x_{48}^2 \lambda_{2458} \lambda_{3458}}.
\end{equation}
where $\lambda_{2458}$ is the Jacobian for the transformation from $x_{6}^{\mu}$ to four propagators related to $x_6$ (which we cut in the above) and $\lambda_{3458}$ is the Jacobian for $x_7$ similarly. For example, 
\begin{equation}\label{eq:lamda2458}
    \lambda_{2458}^{2}=\left|\begin{array}{cccc}
        0 & x_{24}^{2} & x_{25}^{2} & x_{28}^{2} \\
        x_{24}^{2} & 0 & x_{45}^{2} & x_{48}^{2} \\
        x_{25}^{2} & x_{45}^{2} & 0 & x_{58}^{2} \\
        x_{28}^{2} & x_{48}^{2} & x_{58}^{2} & 0
    \end{array}\right|.
\end{equation}
Then, we need to choose cut conditions for the remaining two loops $x_{5}$ and $x_{8}$. We will show two such cuts that are symmetric under the exchange of $x_{2}$ and $x_{3}$ (this is a symmetry of this integral as we can see from the integrand) which can give us elliptic curves. One of them is
\begin{equation}
    {\text{cut: }} x_{15}^{2}, \, x_{25}^{2}, \, \lambda_{2458}, \, \lambda_{3458}, \, x_{38}^{2}, \, x_{18}^{2}
\end{equation}
Under this cut 
\begin{equation}\label{eq:cutcond1}
    \lambda_{2458}=x_{24}^{2}x_{58}^{2}-x_{45}^{2}x_{28}^{2}=0, \, \lambda_{3458}=x_{34}^{2}x_{58}^{2}-x_{35}^{2}x_{48}^{2}=0.
\end{equation}
Then the Jacobian for $x_{5}$ (expressed with $x_{15}^{2}, \, x_{25}^{2}, \, \lambda_{2458}, \, \lambda_{3458}$) can be calculated to be
\begin{equation}
    G_{5}=\sqrt{x_{28}^2 \left(x_{28}^2 x_{34}^2-x_{23}^2 x_{48}^2\right) \left(4 x_{12}^2 x_{13}^2 x_{24}^2 x_{48}^4-x_{14}^4 x_{23}^2 x_{28}^2 x_{48}^2+x_{14}^4 x_{28}^4x_{34}^2\right)}.
\end{equation}
The integrand now becomes (using conditions from \eqref{eq:cutcond1})
\begin{equation}
    I^{\prime}_{168}=\frac{x_{14}^{2}x_{23}^{2}}{\lambda_{1234}G_{5}}\mathrm{d}x_{28}^2\mathrm{d}x_{48}^2 \xrightarrow{\text{renormalize and cut }x_{28}^2} \frac{v}{\lambda_{1234}\sqrt{(u-vX)(4X^2-vX+u)}}\mathrm{d} X.
\end{equation}
The polynomial under square root is homogeneous for $x_{28}^{2}$ and $x_{48}^2$, so we can renormalize with $x_{28}^{2}$ and take $X=x_{12}^{2}x_{48}^{2}/(x_{14}^2x_{28}^{2})$ as the remaining one un-cut variable.
We also define $u=\frac{x_{12}^2x_{34}^2}{x_{13}^2x_{24}^2}=z\bar{z}$,$v=\frac{x_{14}^2x_{23}^2}{x_{13}^2x_{24}^2}=(1-z)(1-\bar{z})$. The integration measure $\mathrm{d} x_{48}^2$ and $\mathrm{d} X$ is recovered in the last step to indicate that we have replaced $x_{48}^{2}$ with $X$. $\lambda_{1234}$, which is defined similarly as $\lambda_{2458}$ in \eqref{eq:lamda2458}, is the Jacobian when we transform from the integration of $x_{8}$ to the integration of $x_{18}^{2},x_{28}^{2},x_{38}^{2},x_{48}^{2}$. Then the elliptic curve is
\begin{equation}
    Y^2=(u-vX)(4X^2-vX+u) 
\end{equation}
Another symmetric cut can be derived from another symmetry of this integral: $x_{2}\leftrightarrow x_{3},x_{6}\leftrightarrow x_{7}$.
\begin{equation}
    {\text{cut: }} x_{15}^{2}, \, x_{35}^{2}, \, \lambda_{2458}, \, \lambda_{3458}, \, x_{28}^{2}, \, x_{18}^{2}
\end{equation}
The resulting elliptic curve will be
\begin{equation}
    Y^2=(1-vX)(4uX^2-vX+1)
\end{equation}
However, this elliptic curve can be transformed to the former one by renormalizing $X$ with $u$, the final remaining integrand is the same as above with a different definition of integration variable $X^{\prime}=x_{12}^{2}x_{34}^{2}x_{48}^{2}/(x_{24}^{2}x_{14}^{2}x_{38}^2)$.
\begin{equation}
    I^{\prime}_{168}=\frac{v}{\sqrt{(u-vX^{\prime})(4{X^{\prime}}^2-vX^{\prime}+u)}}\mathrm{d} X^{\prime}.
\end{equation}
Besides the elliptic cuts, this integral also contains normal cuts. If we choose another cut condition, for example,
\begin{equation}
    {\text{cut: }} x_{15}^{2}, \, x_{25}^{2}, \, x_{35}^2, \, x_{45}^2, \, x_{38}^{2}, \, x_{18}^{2}
\end{equation}
then it can be derived from \eqref{eq:cut168int} that the final leading singularity is
\begin{equation}
    \frac{v}{\lambda_{1234}^2}=\frac{v}{(z-\bar{z})^2}
\end{equation}
where $\lambda_{1234}^2$ comes from the Jacobians of $x_5$ and $x_8$. This is a basic feature in the analysis of leading singularity: when the number of cut conditions is greater than the freedom of loop variable, we need to consider all kinds of choices and different choices may produce very different leading singularities.

We can summarize here that the leading singularities for conformal integrals can be analyzed loop by loop. The subtlety is that we need to consider the cut of Jacobian factors and go over all possible sets of choices of the cut. Usually there are symmetry relations which can reduce the number of cut needed to be analyzed and as we expect, the results are expressions of cross ratios. 
However, above analysis actually relies on the fact that when four cut conditions of a loop variable, e.g. $x_6$, are satisfied, all the other propagators related to it are either solved or the remaining expressions do not depend on this loop at all. For instance, if in the first step of above example, there is an additional $x_{16}^2$ in the denominator, then the analysis can not go on, since we do not know what $x_{16}^2$ is under cut \eqref{eq:cut67}. We will see later how such problems do not appear for conformal integrals by using Gram identities in four dimensions.

\subsubsection{Cut for general vertex in conformal integrals}
\begin{figure}
    \centering
    \includegraphics[width=0.25\textwidth]{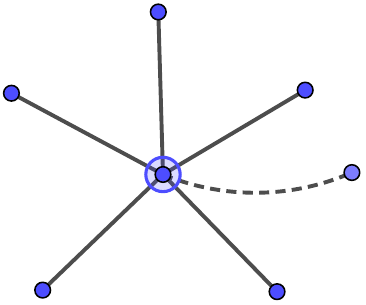}
    \caption{Vertex with a numerator attached.}\label{fig:vertexd5}
\end{figure}

In a conformal integral, if there is a loop vertex appearing in some numerator, then the conformal property implies there will be at least five denominators relevant to this vertex to keep its weight to 4, which we depict in Fig.~\ref{fig:vertexd5}. The idea to calculate the cut of such a vertex is the Gram identity in four dimensions.
Fig.~\ref{fig:vertexd5} involves 7 points, we name them $x_1,x_2,\ldots,x_7$ in order with the central vertex $x_6$ and the point connected to it with dashed line $x_7$. Then the following Gram identity\footnote{We will see such an identity is also important in solving one of the three-loop conformal integrals in Sec.~\ref{sec:threeloopconformal}.} can be constructed
\begin{equation}\label{eq:gram}
    \left|\begin{array}{ccccccc}
        0 & x_{67}^2 & x_{61}^2 & x_{62}^2 & x_{63}^2 & x_{64}^{2} & x_{65}^2 \\
        x_{67}^2 & 0 & x_{71}^2 & x_{72}^2 & x_{73}^2 & x_{74}^{2} & x_{75}^2 \\
        x_{61}^2 & x_{71}^2 & 0 & x_{12}^2 & x_{13}^2 & x_{14}^{2} & x_{15}^2 \\
        x_{62}^2 & x_{72}^2 & x_{12}^2 & 0 & x_{23}^2 & x_{24}^{2} & x_{25}^2 \\
        x_{63}^2 & x_{73}^2 & x_{13}^2 & x_{23}^2 & 0 & x_{34}^{2} & x_{35}^2 \\
        x_{64}^2 & x_{74}^2 & x_{14}^2 & x_{24}^2 & x_{34}^2 & 0 & x_{45}^2 \\
        x_{65}^2 & x_{75}^2 & x_{15}^2 & x_{25}^2 & x_{35}^2 & x_{45}^2 & 0 \\
    \end{array}\right|=0.
\end{equation}
When studying the cut, we can cut any four propagators related to $x_6$ first, for example, $x_{61}^2,x_{62}^2,x_{63}^2,x_{64}^2$. However, there is still $x_{67}^{2}/x_{65}^2$ remaining in the integrand. We need to know what the factor becomes under the cut we choose. The key to solve this problem is to use above Gram identity. It turns into a homogeneous quadratic polynomial in $x_{67}^2,x_{65}^2$. Therefore $x_{67}^2/x_{65}^2$ is solved by expression of other vertices. Then we can continue our analysis of the cut for remaining loop variables. This argument can be directly generalized to vertices with higher degree (more than one numerator attached). 
Let us take the following three-loop integral as an example which is depicted as $I^{(3)}_{15}$ in Fig.~\ref{fig:threeloopbasis}.
The integrand takes the following form 
\begin{equation}
    I_{15}^{(3)}=\frac{x_{17}^2 x_{23}^2 x_{24}^2 x_{34}^2 x_{56}^2}{x_{15}^2 x_{16}^2 x_{25}^2 x_{26}^2 x_{27}^2 x_{35}^2 x_{36}^2 x_{37}^2 x_{45}^2 x_{46}^2 x_{47}^2 x_{57}^2 x_{67}^2}.
\end{equation}
It is special in the sense that all its loop vertices have degree 5 in the denominator, that is, five cut conditions for each loop in the denominator. When we choose any four of them, the remaining one must be solved. This is indeed the case after we apply the Gram identity. We take two different typical cuts as examples.
The first case is
\begin{equation}\label{eq:cutI151}
    {\text{cut: }} x_{51}^2, x_{52}^2, x_{53}^2, x_{54}^2, x_{61}^2, x_{62}^2, x_{63}^2, x_{64}^2, x_{27}^2, x_{37}^2, x_{47}^2.
\end{equation}
Under this cut, the Gram identity gives one nontrivial constraint
\begin{equation}
    \frac{x_{56}^2}{x_{57}^2x_{67}^2}=-\frac{\lambda_{1234}^2}{x_{23}^2x_{24}^2x_{34}^2x_{17}^4}.
\end{equation}
It reduces the remaining expressions and after cancellation we get the last cut condition for $x_{7}$, $x_{17}^2=0$. Then the leading singularity is
\begin{equation}
    -\frac{1}{\lambda_{1234}}=-\frac{1}{z-\bar{z}}.
\end{equation}
The last cut condition only emerges after we impose the earlier cut conditions\footnote{The reader may wonder how one can determine when we should impose a set of conditions first to generate new conditions. In fact it can be systematically tackled in the following way: we impose 12 conditions for three-loop vertices first. For example, in \eqref{eq:cutI151}, we supplement an additional cut condition $x_{67}^2=0$. Then the Gram identity requires $x^{2}_{17}=0$ or $x^{2}_{56}=0$, which means we will come across a $0/0$ indeterminate under this full cut. Then we should relax one condition and consider the limit. There are different ways to relax the cut, we should consider all possible cases and they may give the same or different results.}.

The second case is
\begin{equation}\label{eq:cutI152}
    {\text{cut: }} x_{61}^2, x_{62}^2, x_{63}^2, x_{64}^2, x_{57}^2, x_{67}^2, x_{27}^2, x_{37}^2.
\end{equation}
The Gram identity under this cut will give a nontrivial identity
\begin{equation}\label{eq:ration1747}
    \frac{x_{17}^{2}}{x_{47}^2}=\frac{x_{13}^2x_{24}^2+x_{12}^2x_{34}^2-x_{14}^2x_{23}^2\pm \lambda_{1234}}{2x_{24}^2x_{34}^2}
\end{equation}
Then we find an interesting thing, on the support of this cut, this integral is equivalent to the three-loop hard integral that has been studied before in~\cite{Drummond:2013nda} up to some constant normalization factor. Here we give the integrand of hard integral\footnote{This differs from the standard definition with a permutation of external legs, but this does not affect our discussion.}
\begin{equation}
    I^{(3)}_{12}=\frac{x_{12}^2 x_{23}^2 x_{34}^2 x_{56}^2}{x_{15}^2 x_{16}^2 x_{25}^2 x_{26}^2 x_{27}^2 x_{35}^2 x_{36}^2 x_{37}^2 x_{45}^2 x_{46}^2 x_{57}^2 x_{67}^2}.
\end{equation}
And the constant factor can be obtained from \eqref{eq:ration1747} to be $(1+u-v\pm \lambda_{1234})/(2u)$. So this integral will inherit leading singularities from the hard integral, deformed by a normalization factor.

Though we find that Gram identities are necessary if we want to perform leading singularities for certain conformal integrals like the above case, the usage of Gram identity is clumsy when there are too many different leading singularities. Since our aim is to calculate the integral analytically, we can directly use the Gram identities to generate integral identities like integration by parts. We will show that $I^{(3)}_{15}$ can be solved by reducing to other integrals that are known using the Gram identity later\footnote{The idea of using Gram identities to simplify the calculation of periods has been exploited in~\cite{Zhang:2024ypu}, here we use the same idea to simplify the calculation of some nonplanar conformal integrals.}. 

\subsubsection{Construction of integrals with single leading singularity}\label{sec:splitls}
\begin{figure}[htbp]
    \centering
    \includegraphics[width=0.3\textwidth]{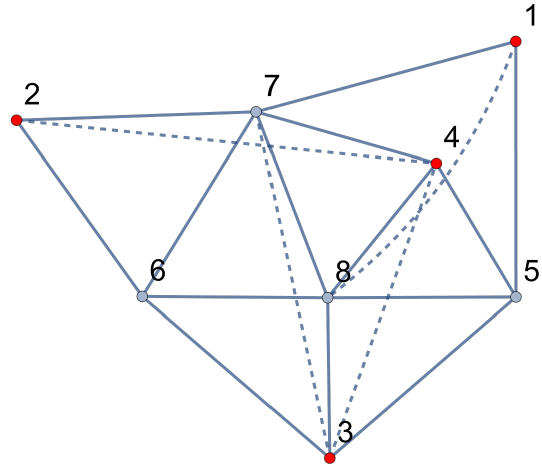}
    \caption{The conformal integral which is defined to be one of two easy types at four loop, due to the same leading singularities with three-loop easy integrals.}\label{fig:I4L176}
\end{figure}

Similar to what has been done in~\cite{He:2025rza,Kuo:2025bdr}, where the canonical integrands have been constructed for the four-loop four-point correlator in $\mathcal{N}=4$ SYM. In this section, we exploit the information in the process of leading singularity analysis to split a conformal integral into several parts with single leading singularities, which facilitates our bootstrap of each part separately. Let us show this with a four-loop conformal integral $I^{(4)}_{176}$ which we will bootstrap later.
The integrand of $\mathcal{I}^{(4)}_{176}$ is
\begin{equation}
    I_{176}^{(4)}=\frac{x_{18}^2 x_{34}^2 x_{24}^2 x_{37}^2}{x_{15}^2 x_{17}^2 x_{26}^2 x_{27}^2 x_{35}^2 x_{36}^2 x_{38}^2 x_{45}^2 x_{47}^2x_{48}^2 x_{58}^2 x_{67}^2 x_{68}^2 x_{78}^2}.
\end{equation}
which is depicted in Fig.~\ref{fig:I4L176}.
We can first cut out $x_5$ and $x_6$ and arrive at a leading singularity
\begin{equation}
    I^{\prime}_{176}=\frac{x_{18}^2 x_{34}^2 x_{24}^2 x_{37}^2}{x_{17}^2 x_{27}^2 x_{38}^2 x_{47}^2x_{48}^2x_{78}^2\lambda_{1348}\lambda_{2378}}
\end{equation}
where $\lambda_{1348}$ and $\lambda_{2378}$ simplifies to
\begin{equation}
\begin{aligned}
    &\lambda_{1348} \to x_{18}^{2}x_{34}^{2}-x_{13}^{2}x_{48}^{2}, \\
    &\lambda_{2378} \to x^{2}_{37}x^{2}_{28}
\end{aligned}
\end{equation}
under the support of cut $x_{38}^{2}=x^{2}_{78}=0$. $x_{18}^{2}x_{34}^{2}-x_{13}^{2}x_{48}^{2}$ in the denominator is where $1/(z-\bar{z})/(1-u)$ originates. So the easiest way to remove this leading singularity is to apply the following replacement to the original integrand: $x^{2}_{18}x^{2}_{34}\to x^{2}_{18}x^{2}_{34}-x^{2}_{13}x^{2}_{48}$, which should cancel the pole originating from $\lambda_{1348}$. In other words, we can split the original integrand into the following two parts:
\begin{equation}
\begin{aligned}
    &I_{176}=\frac{(x_{18}^2 x_{34}^2-x^{2}_{13}x^{2}_{48}+x^{2}_{13}x^{2}_{48}) x_{24}^2 x_{37}^2}{x_{15}^2 x_{17}^2 x_{26}^2 x_{27}^2 x_{35}^2 x_{36}^2 x_{38}^2 x_{45}^2 x_{47}^2x_{48}^2 x_{58}^2 x_{67}^2 x_{68}^2 x_{78}^2}= I^{a}_{176} + I^{b}_{176}=\\ 
    &\frac{(x_{18}^2 x_{34}^2-x^{2}_{13}x^{2}_{48}) x_{24}^2 x_{37}^2}{x_{15}^2 x_{17}^2 x_{26}^2 x_{27}^2 x_{35}^2 x_{36}^2 x_{38}^2 x_{45}^2 x_{47}^2x_{48}^2 x_{58}^2 x_{67}^2 x_{68}^2 x_{78}^2}
    \!+\!\frac{x^{2}_{13} x_{24}^2 x_{37}^2}{x_{15}^2 x_{17}^2 x_{26}^2 x_{27}^2 x_{35}^2 x_{36}^2 x_{38}^2 x_{45}^2 x_{47}^2x_{58}^2 x_{67}^2 x_{68}^2 x_{78}^2} .
\end{aligned}
\end{equation}
The first part $I^{a}_{176}$ is now free of $1/(z-\bar{z})/(1-u)$ and it has only one leading singularity $1/(z-\bar{z})$. The same leading singularity analysis for the second part $I^{b}_{176}$ shows that it has only one other leading singularity $1/(z-\bar{z})/(1-u)$. In such a way, we have split the contributions from two different leading singularities. Such a split is not unique, since we can always add and subtract some additional parts from the original integrals as long as it keeps the leading singularity of each single part unique. For example, we can split the original integral into three parts with each part a single unique leading singularity:
\begin{equation}\label{eq:expI176}
\begin{aligned}
    &I_{176}=\frac{(x_{18}^2 x_{34}^2-x^{2}_{13}x^{2}_{48}-\alpha x^{2}_{14}x^{2}_{38}) x_{24}^2 x_{37}^2}{x_{15}^2 x_{17}^2 x_{26}^2 x_{27}^2 x_{35}^2 x_{36}^2 x_{38}^2 x_{45}^2 x_{47}^2x_{48}^2 x_{58}^2 x_{67}^2 x_{68}^2 x_{78}^2}+ \\ 
    &\frac{x^{2}_{13} x_{24}^2 x_{37}^2}{x_{15}^2 x_{17}^2 x_{26}^2 x_{27}^2 x_{35}^2 x_{36}^2 x_{38}^2 x_{45}^2 x_{47}^2x_{58}^2 x_{67}^2 x_{68}^2 x_{78}^2} \!+\! \frac{\alpha x^{2}_{14} x_{24}^2 x_{37}^2}{x_{15}^2 x_{17}^2 x_{26}^2 x_{27}^2 x_{35}^2 x_{36}^2 x_{45}^2 x_{47}^2 x_{48}^2 x_{58}^2 x_{67}^2 x_{68}^2 x_{78}^2}.
\end{aligned}
\end{equation}
The new third part will be denoted as $I^{c}_{176}$. These three parts have single leading singularities $1/(z-\bar{z})$, $1/(z-\bar{z})/(1-u)$ and $1/(z-\bar{z})$ respectively. We can find that the numerators we add to cancel some leading singularities are Jacobians (which is a Gram determinant) like the above $\lambda_{1348}$. They usually take the form of combination of $x^{2}_{ab}x^{2}_{cd}$, $x^{2}_{ac}x^{2}_{bd}$ and $x^{2}_{ad}x^{2}_{bc}$. 

However, we need to note that this construction of integral with single leading singularity is not easy for all kinds of conformal integrals. It is suitable for integrals with leading singularity that is rational functions of u and v, just like $1/(1-u)$. Finally, we are ready to bootstrap each single part as long as we know the function space following each leading singularity.

\subsubsection{Single-valued multiple polylogarithms and the ansatz construction}\label{sec:ansatz}
For conformal integrals which are free from elliptic cut, we expect them to evaluate to single-valued multiple polylogarithms (svMPLs)~\cite{Drummond:2012bg,Dixon:2012yy,Chavez:2012kn,Schnetz:2013hqa,Schnetz:2021ebf}. Here we briefly review this function space.
We adopt the following definition which is also used within the package \texttt{HyperlogProcedures}. The single-valued harmonic polylogarithms (svHPLs) $\mathrm{I}_{z,...,0}$ are defined recursively as follows:
\begin{equation}\label{eq:svhpldef}
    \begin{aligned}
    &\mathrm{I}_{z,a_{n},a_{n-1},\ldots,a_{1},0}=\int_{sv}\frac{\mathrm{d}z}{z-a_{n}}\mathrm{I}_{z,a_{n-1},\ldots,a_{1},0}, \, a_{n}=0,1; \\
    &\mathrm{I}_{z,0}=1; \mathrm{I}_{z,0,0}=\log z\bar{z}, \, \mathrm{I}_{z,1,0}=\log(1-z)(1-\bar{z}).
    \end{aligned}
\end{equation}
$\int_{sv}$ is the single-valued integration defined in \cite{Schnetz:2013hqa}, which not only is an integration of $z$, but also accounts for the antiholomorphic part in $\bar{z}$ and the boundary condition which requires the integration vanishes when $z$ approaches 0 (regularized around $z=0$). It keeps the result single-valued without branch cuts along the real axis. 
The integration kernel in \eqref{eq:svhpldef} can be further generalized such that $a_n$ can be expressions of M\"obius transformation of $\bar{z}$, such as $\bar{z}$ or $1/\bar{z}$. The single-valued integration is also generalized correspondingly~\cite{Schnetz:2021ebf}. In this paper, we call such generalized integrals single-valued multiple polylogarithms (svMPLs). 
The shorthand $\mathcal{L}_{a_1,a_2,\ldots,a_n}$ is used to denote $\mathrm{I}_{z,a_{n},a_{n-1},\ldots,a_{1},0}$ defined above. We will also call $a_n$ the letter and $\{a_1,a_{2},\ldots,a_n\}$ the word for svMPLs. This letter should be distinguished from the standard notation for symbol letters which are actually $z-a_n$ and $\bar{z}-a_n$.

\begin{figure}
    \centering
    \includegraphics[width=0.3\textwidth]{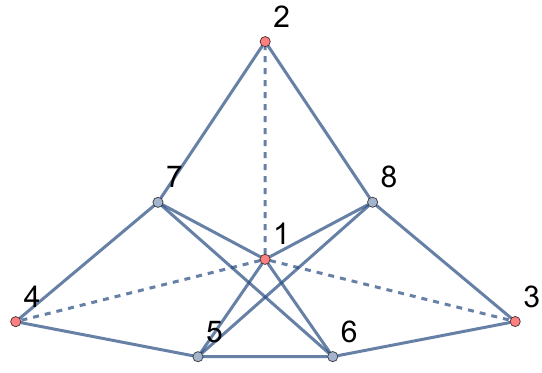}
    \caption{A four-loop conformal integral with a new letter $\bar{z}$ in the function space. Red points denote four external points.}\label{fig:I422}
\end{figure}

To bootstrap the analytic results of fixed-loop conformal integrals, we need more information to construct a proper function space. The basic information is the possible set of letters for svMPLs involved. It is better if we also know the possible positions of these letters in the sequence.
Another important information is the possible weight of svMPLs. Especially, we want to know whether this conformal integral is uniform weight and free from weight drop ($L$-loop conformal integrals are expected to be of highest weight $2L$). We call a conformal integral \textit{rigid} if it is of uniform weight and free of weight drop, while it is allowed to have multiple leading singularities. Usually, leading singularity analysis can give such information as well. For example, if we find such a cut of the integrand that results in a double pole of remaining integration variables, it usually indicates a weight drop for the integral because there is no $\mathrm{d}\log$ integration on the support of this cut. 
However, even if the leading singularity analysis indicates the rigidity of the conformal integral, that is, we know the weight of the functions, how can we know the possible letters involved? It is very tricky how the letters appearing in the svMPLs are related to the leading singularities alone. There are examples that the letters in function space can not be probed by leading singularity analysis. For example,
\begin{equation}
    I^{(4)}_{22}=\frac{x_{12}^2 x_{13}^2 x_{14}^2}{x_{15}^2 x_{16}^2 x_{17}^2 x_{18}^2 x_{27}^2 x_{28}^2 x_{36}^2 x_{38}^2 x_{45}^2 x_{47}^2 x_{56}^2 x_{58}^2 x_{67}^2}.
\end{equation}
It is depicted in Fig.~\ref{fig:I422}. The leading singularity analysis shows that it is rigid and has only one leading singularity $1/(z-\bar{z})$. However, the function space for this integral involves a new letter $\bar{z}$ which is absent from the standard leading singularity analysis.
Here we propose to use generalized boxing to detect possible hidden letters for a conformal integral like above. We call it a generalized boxing because for standard boxing~\cite{Drummond:2006rz,Drummond:2012bg}, we need an external vertex which connects to the internal (loop) vertex through a single edge. In such standard case, the right hand of the boxing differential equation is a single lower-loop conformal integral\footnote{For an $L$-loop conformal integral $\Phi^{(L)}\equiv f^{(L)}/(z-\bar{z})$ satisfying boxing differential equation, we will have $z\bar{z}\partial_{z}\partial_{\bar{z}}f^{(L)}=cf^{(L-1)}$ or $(1-z)(1-\bar{z})\partial_{z}\partial_{\bar{z}}f^{(L)}=cf^{(L-1)}$ depending on the normalization, where $c$ is some constant.}. However, for generalized boxing, we do not require such a single edge connection. In this case, the right hand of the boxing differential equation will be a sum of several conformal integrals (see App.~\ref{app:boxing} for the derivation). They are hard to solve, but we only care about the leading singularities of the lower-loop integrals derived from boxing differential equation. Since they indicate the possible letters that will appear in the last two entries of svMPLs\footnote{The solving of boxing differential equation of conformal integral is the single-valued integration of $z,\bar{z}$, so we can imagine that if there is a pole $1/(z-\bar{z})$ in $f^{(L-1)}$ (that is, a double pole $1/(z-\bar{z})^2$ in conformal integral $\Phi^{(L)}$), it will result in the letter $\bar{z}$ in svMPLs.}. This procedure can be repeated so that we can detect possible new letters in deeper depth of the word for svMPLs.

Let us still take $I^{(4)}_{22}$ as the example. We can apply the boxing to external vertex $x_4$. This will result in three four-loop conformal integrals with double poles along with two lower-loop conformal integrals. We concentrate on these two lower-loop ones. One corresponds to identifying $x_5$ with $x_4$ and another corresponds to identifying $x_7$ with $x_4$.
For the first case, the integral splits into a direct product of a one-loop integral (with loop vertex $x_8$) and a two-loop integral (with loop vertex $x_6$ and $x_7$). The one-loop integral is a box with leading singularity $1/(z-\bar{z})$ and the two-loop integral is double box with the same leading singularity $1/(z-\bar{z})$. Thus this three-loop integral has leading singularity $1/(z-\bar{z})^2$, which implies that the original four-loop integral $I^{(4)}_{22}$ may have a new letter $\bar{z}$ in the last two entries of svMPLs.
For the second case, the integral is a genuine three-loop integral with leading singularity $u/(z-\bar{z})$. At the same time, we can also apply the boxing to external vertex $x_2$. This will result in two three-loop integrals as well. However, we can act the boxing on these two three-loop integrals again, which results in the product of two boxes, thus with leading singularity $1/(z-\bar{z})^2$. Then we can infer that the letter $\bar{z}$ can also appear in the last four entries. Since $I^{(4)}_{22}$ can be calculated by \texttt{HyperlogProcedures}, the final result agrees with the above analysis from boxing.

Although the boxing procedure probes possible new letters well for some conformal integrals with leading singularity $1/(z-\bar{z})$, it becomes subtle for integrals with singularities other than $1/(z-\bar{z})$. We take $I^{(4)}_{176}$ studied in the last section which can not be calculated by \texttt{HyperlogProcedures} as an example. First of all, $I^{(4)}_{176}$ does not involve any elliptic cut and its leading singularity is $1/(z-\bar{z})$ and $1/(1-u)/(z-\bar{z})$.
Naively the ansatz can be written as 
\begin{equation}
    \mathcal{I}_{176}^{(4)}=\sum_{i}c_i\frac{\mathcal{L}^{\text{odd}}_{\text{ans},i}}{z-\bar{z}} + \sum_{j}d_j\frac{\mathcal{L}^{\text{odd}}_{\text{ans},j}}{(z-\bar{z})(1-z\bar{z})}, \quad \left.\mathcal{L}^{\text{odd}}_{i}\right|_{z\leftrightarrow\bar{z}}=-\mathcal{L}^{\text{odd}}_{i} 
\end{equation}
The parity even property under exchange $z\leftrightarrow\bar{z}$ of the conformal integral (thus the parity odd property of our function space with leading singularity $1/(z-\bar{z})$) will give rather strong constraints. 
And we note that the splitting of $\mathcal{I}_{176}^{(4)}$ into three parts with single leading singularities (in \eqref{eq:expI176}) gives us more constraints than a single ansatz as above.
So in our calculation, we use the following construction instead.
\begin{equation}\label{eq:I4L176abc}
    \mathcal{I}_{176}^{a}=\sum_{i}c_i\frac{\mathcal{L}^{\text{odd}}_{\text{ans},i}}{z-\bar{z}}, \quad \mathcal{I}_{176}^{b} = \sum_{j}d_j\frac{\mathcal{L}^{\text{odd}}_{\text{ans},j}}{(z-\bar{z})(1-z\bar{z})},\quad \mathcal{I}_{176}^{c} = \sum_{k}e_k\frac{\mathcal{L}^{\text{odd}}_{\text{ans},k}}{z-\bar{z}}.
\end{equation}
$\mathcal{I}_{176}^{c}$ is not necessary to obtain $\mathcal{I}_{176}^{(4)}$ if we take $\alpha=0$ in \eqref{eq:ansatzI176}. However, it is still interesting that we can obtain these three different integrals at the same time. To further constrain our ansatz, we need to analyze what the function space $\mathcal{L}^{\text{odd}}_{\text{ans},i}$ should be used for above three integrals. 
In such a case, the boxing is no longer suitable. First, the integrand $I^{a}_{176}$ is composed of several parts, so the action of boxing on its integrand is not very illuminating. Second, the new leading singularity $1/(1-z\bar{z})$ in $\mathcal{I}^{b}_{176}$ will be inherited by the lower-loop integrals if we apply the boxing differential operator $z\bar{z}\partial_{z}\partial_{\bar{z}}$ by definition, which obscures the information we may extract for $\mathcal{L}^{\text{odd}}_{\text{ans},j}$. Finally, the boxing for $\mathcal{I}^{c}_{176}$ will introduce double poles in the lower-loop integrals, which brings challenges to its leading singularity analysis.
All these form an obstruction to using the boxing procedure as above. However, we observe from some similar integrals an empirical rule (which is presented in Appendix~\ref{app:reducerule}) to set up the following ansatz for these three integrals
\begin{equation}\label{eq:ansatzI176}
        \mathcal{L}_{\text{ans}}: \quad \mathcal{L}_{\frac{1}{\bar{z}},\frac{1}{\bar{z}},\ldots}, \,
        \mathcal{L}_{0,\frac{1}{\bar{z}},\ldots},\, \mathcal{L}_{1,\frac{1}{\bar{z}},\ldots},
        \mathcal{L}_{\frac{1}{\bar{z}},0,\ldots},\,
        \mathcal{L}_{\frac{1}{\bar{z}},1,\ldots},\,\mathcal{L}_{\ldots}
    \end{equation}
where $\ldots$ denotes only sequences of $0$ and $1$. For such ansatz \eqref{eq:ansatzI176} after imposing parity condition, it turns out that up to weight 8, there are only 148 of them, among which 135 are simply svHPLs with only $0$ and $1$ as letters.

Generally, for integrals with new poles like $1/(1-z\bar{z})$ in leading singularities, we will include this new pole to our candidate for letters, though it may be spurious and does not appear in the final results. The empirical rule in Appendix~\ref{app:reducerule} will be more accurate but it is still totally an observation. The ansatz is justified by the success of bootstrap and the verification of numerical values.

\subsection{Boundary Conditions}\label{sec:boundary}
\begin{figure}[htbp]
    \centering
    \includegraphics[width=0.5\textwidth]{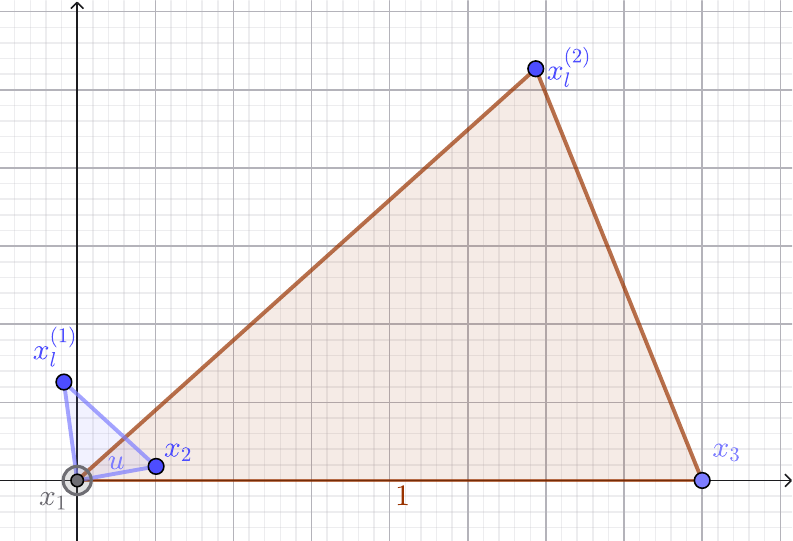}
    \caption{The two different regions for loop vertex in different colors. In each region, we can only see one triangle, the other triangle either shrunk to invisible, or expanded to the whole plane, thus each with one scale. The loop vertices are divided into two sets.}\label{fig:region}
\end{figure}

All the constraints for the ansatz we construct come from the boundary condition\footnote{Note that we can always permute the external legs of conformal integrals, so there are actually six inequivalent boundaries.} $x_2\to x_1$ of the conformal integrals along with discrete symmetry (if the integral has additional symmetries). We can fix $x_1=0$, $x_3=1$ and $x_4=\infty$. So this limit is simply $u\to 0, v\to 1$. We follow the method in \cite{Chicherin:2018avq} and perform the expansion by region for this integral to get the boundary data, which can be done in the integrand level. In this kinematic boundary, there are two small scales $u$ and $Y=1-v$, and for each vertex $x_{l}$ to be integrated, it has only two regions, one is close to $x_1=0$, $(x_{l}-x_{1})^2$ comparable to $u\ll 1$ and $(x_{l}-x_{3})^2$ close to 1, so
\begin{equation}\label{eq:series1}
    \frac{1}{x_{l3}^2}=\frac{1}{1-(1-x_{l3}^2)}=\sum_{n=0}^{N}(1-x_{l3}^2)^n=\sum_{n=0}^{N}(2x_l\cdot x_3-x_{l1}^2)^n.
\end{equation}
The other is away from $x_1=0$, and $(x_{l}-x_{2})^2- (x_{l}-x_{1})^2$ is comparable to $u\ll 1$, so we should perform the expansion
\begin{equation}\label{eq:series2}
    \frac{1}{x_{l2}^2}=\frac{1}{x_{l1}^2-(x_{l1}^2-x_{l2}^2)}=\sum_{n=0}^{N}\frac{(x_{l1}^2-x_{l2}^2)^n}{(x_{l1}^2)^{n+1}}=\sum_{n=0}^{N}\frac{(2x_{l}\cdot x_2-x_{12}^2)^n}{(x_{l1}^2)^{n+1}}.
\end{equation}
We can either perform expansion for $1/x^2_{l2}$ or $1/x^2_{l1}$, here we choose $1/x^2_{l2}$. 
At last, for two loops living in different regions
\begin{equation}\label{eq:series3}
    \frac{1}{x_{ab}^2}=\frac{1}{x_{b1}^2-(x_{b1}^2-x_{a1}^2)}=\sum_{n=0}^{N}\frac{(2x_{a}\cdot x_b-x_{a1}^2)^n}{(x_{b1}^2)^{n+1}}.
\end{equation}
where $b$ denotes the vertex away from $x_1$. In such a way, we disentangle two loops in different regions.
The regions for the vertices are displayed in Fig.~\ref{fig:region}, two regions correspond to two triangles with different colors.
The next thing is to truncate the above series expansion to a given order. Note that each numerator in \eqref{eq:series1}, \eqref{eq:series2} and \eqref{eq:series3} is of order $Y^n$. For three-loop integrals, we only keep up to $\mathcal{O}(Y^3)$ and for four-loop integrals, we keep up to $\mathcal{O}(Y^4)$.

\begin{figure}[htbp]
    \centering
    \raisebox{-0.5\height}{\includegraphics[scale=0.25]{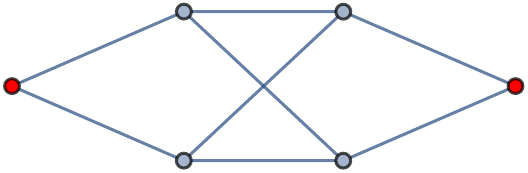}}\quad
    \raisebox{-0.5\height}{\includegraphics[scale=0.25]{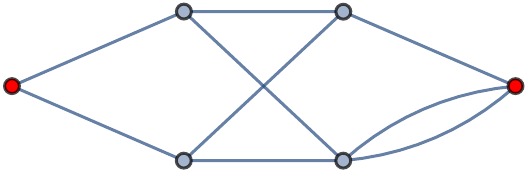}}\quad
    \raisebox{-0.5\height}{\includegraphics[scale=0.25]{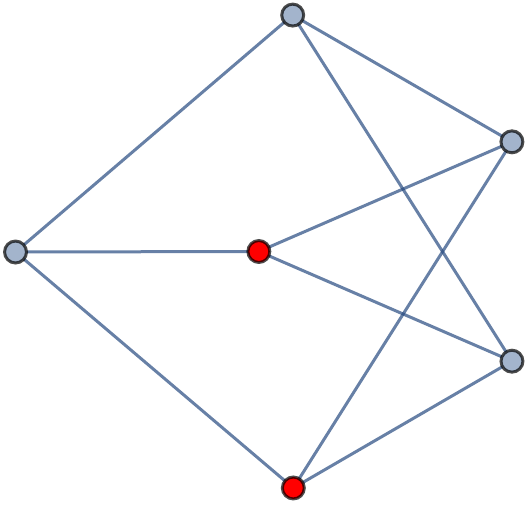}}\quad
    \raisebox{-0.5\height}{\includegraphics[scale=0.25]{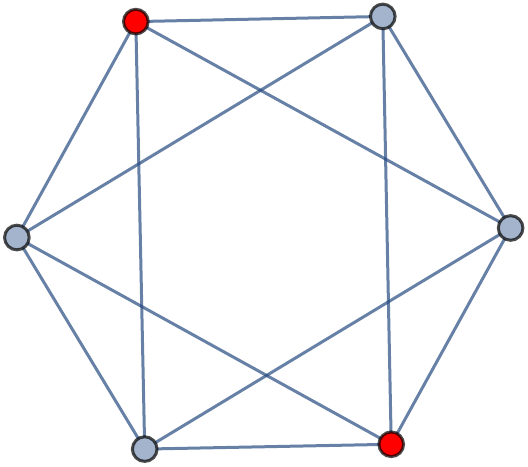}}
    \caption{The four nonplanar master integrals. Red points are two external points.}
    \label{fig:nonplanar}
\end{figure}

After performing tensor reduction, the boundary is reduced to a set of (two-point) $p$-integrals with scales either $x_{12}^{2}=u$ or $x_{13}^{2}=1$. After IBP reduction by \texttt{LiteRed2}~\cite{Lee:2013mka} and applying four-loop master integral results for $p$-integrals~\cite{Baikov:2010hf,Lee:2011jt}, we only keep the final expression up to $\mathcal{O}(\epsilon^{0})$. 
We note here that since our integrals are in dual coordinate space, they are vertex integrals rather than loop integrals\footnote{However, we will sometimes call these vertices loop variables, because they play a role similar to loop momenta. And in the planar case, they can indeed be taken as loop momenta.}. Only planar ones can be mapped to loop integrals and they can be really taken as loop momenta. At three loops, there are totally 5 master integrals, and they can all be mapped to loop integrals. At four loops, there are totally 24 masters for these two-point integrals. 20 of them can be mapped to loop integrals, 4 of them are nonplanar in vertex coordinate space. Since we do not constrain ourselves to the evaluation of planar vertex integrals, we need to work out the 4 nonplanar two-point integrals. Fortunately, they can all be evaluated directly in \texttt{HyperlogProcedures}. We give the definition of these 4 masters and their results here.
The topology is defined as
\begin{equation}
    \begin{aligned}
       &l_{5}^2, (l_5-p)^2, l_6^2, (l_6-p)^2, l_7^2, (l_7-p)^2, l_8^2, (l_8-p)^2, (l_5-l_6)^2, (l_{5}-l_7)^2, (l_5-l_8)^2, \\
       &(l_6-l_7)^2, (l_6-l_8)^2, (l_7-l_8)^2.
    \end{aligned}
\end{equation}
where $p^2=u$ or $p^2=1$. The four nonplanar masters are chosen by \texttt{LiteRed2} to be
\begin{equation}
    \begin{aligned}
        &N_1\equiv G[0, 1, 0, 1, 1, 0, 1, 0, 0, 1, 1, 1, 1, 0], N_2\equiv G[{0, 2, 0, 1, 1, 0, 1, 0, 0, 1, 1, 1, 1, 0}], \\
        &N_3\equiv G[0, 0, 1, 1, 1, 1, 1, 1, 1, 1, 1, 0, 0, 0], N_4\equiv G[1, 1, 1, 1, 1, 1, 1, 1, 0, 1, 1, 1, 1, 0].
    \end{aligned}
\end{equation}
They can be easily drawn if we take $l_5,l_6,l_7,l_8$ as coordinates in Fig.~\ref{fig:nonplanar}, rather than loop momenta.
Using the function \texttt{TwoPoint} with the edge weight set to $1/(-1+d/2)$\footnote{This is a convention in package \texttt{HyperlogProcedures}, the edge weight has been deformed to $d/2-1$, so we must cancel it. We thank Oliver Schnetz for informing us such a convention. Also note that in the package, $d=4-\varepsilon$. However, in our context, we use $d=4-2\epsilon$ as usually defined.} in the package \texttt{HyperlogProcedures}, we finally obtain the results\footnote{We have also checked results of some known planar masters with the same procedure, they match perfectly.}
\begin{equation}
    \begin{aligned}
        N_1=&u^{-4\epsilon}\left[\frac{5 \zeta_5}{\epsilon}-13 \zeta _3^2+35 \zeta _5+\frac{5 \pi ^6}{378}\right.\\
        &\hspace{45pt}\left.+\epsilon\left(-91 \zeta _3^2-\frac{13 \pi ^4 \zeta _3}{30}-\frac{5 \pi ^2 \zeta _5}{3}+195 \zeta _5+\frac{345 \zeta _7}{4}+\frac{5 \pi ^6}{54}\right)+\mathcal{O}(\epsilon^2)\right],\\
        N_2=&u^{-1-4\epsilon}\left[-\frac{20 \zeta_5}{\epsilon}-\frac{10 \pi ^6}{189}-8 \zeta _3^2-40 \zeta _5\right.\\
        &\hspace{45pt}\left.+\epsilon\left(-\frac{20 \pi ^6}{189}-\frac{4 \pi ^4 \zeta _3}{15}-16 \zeta _3^2-80 \zeta _5+\frac{20}{3} \pi ^2 \zeta _5-520 \zeta _7\right)+\mathcal{O}(\epsilon^2)\right],\\
        N_3=&u^{-1-4\epsilon}\left[36 \zeta _3^2+\epsilon\left(\frac{6 \pi ^4 \zeta _3}{5}+360 \zeta _3^2-567 \zeta _7\right)+\mathcal{O}(\epsilon^2)\right],\\
        N_4=&u^{-4-4\epsilon}\left[-\frac{10 \zeta_5}{\epsilon}-\frac{5 \pi ^6}{189}+2 \zeta _3^2-30 \zeta _5\right.\\
        &\hspace{45pt}\left.+\epsilon\left(-\frac{5 \pi ^6}{63}+\frac{\pi ^4 \zeta _3}{15}+6 \zeta _3^2+90 \zeta _5+\frac{10}{3} \pi ^2 \zeta _5+\frac{201 \zeta _7}{2}\right)+\mathcal{O}(\epsilon^2)\right].
    \end{aligned}
\end{equation}
In the end, the boundary conditions will be evaluated to expressions in $\log u$ and $Y$ since we only keep the leading order of $u$.

Finally, we comment on several technical points. First, all conformal integrals are invariant under the following permutations
\begin{equation}
    \{1\leftrightarrow 2, 3\leftrightarrow 4\}, \{1\leftrightarrow 3, 2\leftrightarrow 4\}, \, \{1\leftrightarrow 4, 2\leftrightarrow 3\}
\end{equation}
because the cross ratios $u,v$ are invariant under the above permutations. However, the integrands are not necessarily invariant under above permutations, thus the difficulty of region expansion in the integrand level will be different across equivalent expressions. In our program, we choose from several equivalent integrands which generate least terms in the process after series expansions. This may greatly reduce some intermediate results.
Second, we calculate the boundary conditions for the remaining 6 non-equivalent permutations of external points $x_1,x_2,x_3$ to get most constraints. 
A subtle point here is, for some integrals like $I^{b}_{176}$, there is a kinematic factor $x_{13}^{2}x_{24}^{2}$ in the numerator which will be proportional to $x_{12}^{2}=u$ under the permutation of $2\leftrightarrow 3$. In such cases, its boundary condition starts from $\mathcal{O}(u^1)$ rather than $\mathcal{O}(u^0)$. So to get all possible conditions, we will strip the constant factors that only depend on external points when calculating the boundary conditions for an integral and compensate for its effect later.

The calculation of the boundary conditions of conformal integrals of three-loop and four-loop can be performed independently by the boundary agent in the package we provide. It requires the installation of Mathematica package \texttt{LiteRed2}. If the package \texttt{FiniteFlow}~\cite{Peraro:2019svx} is also installed, it may perform better though this is not necessary.
\subsection{Series Expansions of ansatz}\label{sec:series}
Now we discuss the series expansions for the ansatz constructed. The ansatz is composed of two parts: leading singularities and the single-valued MPLs following them. Since our boundary is calculated to expressions of $\log u$ and $Y$, we must series expand our ansatz to expressions of them as well. However, our functions are all single-valued MPLs written with $z$ and $\bar{z}$, so we must do some variable transformation. This seems trivial but there are several subtleties here.

First, the boundary limit $u\to 0,Y\to 0$ is not equivalent to the limit of $\bar{z}\to 0, z\to 0$ because of the subtlety of the multivariate limit. In the calculation of boundary conditions in the last section, we first take $u\to0$, then $Y\to 0$ up to a certain order. However, the limit we will take for single-valued MPLs is first $\bar{z}\to 0$, then $z\to 0$. This matters because the series expansion of leading singularities $1/(z-\bar{z})$, $1/(z-\bar{z})^2$ indeed depend on this order, though the final physical result does not. The subtlety is that we can derive from the order $\bar{z}\to0, z\to0$ that $u\to 0, Y\to 0$, since $u=z\bar{z}, Y=z+\bar{z}-z\bar{z}$. But we can not derive in the reverse order. The fundamental reason is certainly due to that the configuration space of $z,\bar{z}$ is a double cover of configuration space of $u,Y$. We summarize the relation between these two limits and branch choice for $z,\bar{z}$ in Table~\ref{tab:map}.
\begin{table}[htbp]
    \centering
    \begin{tabular}{c|c|c}
    \hline\hline
     limit in $u,v$    &  limit in $z,\bar{z}$ & expressions of $z,\bar{z}$ \\ \hline 
       \multirow{2}{*}{$u\to 0, Y\to 0^{+}$}
         &  $\bar{z}\to 0, z\to 0$ &
       {\footnotesize 
           $z=\frac{1+u-v+\sqrt{-4u+(1+u-v)^2}}{2}$,
           $\bar{z}=\frac{1+u-v-\sqrt{-4u+(1+u-v)^2}}{2}$
       } \\[6pt]
       &  $z\to 0, \bar{z}\to 0$ & {\footnotesize 
           $z=\frac{1+u-v-\sqrt{-4u+(1+u-v)^2}}{2}$,
           $\bar{z}=\frac{1+u-v+\sqrt{-4u+(1+u-v)^2}}{2}$
       } \\
       \hline
       \multirow{2}{*}{$u\to 0, Y\to 0^{-}$}
         &  $z\to 0, \bar{z}\to 0$ & {\footnotesize 
           $z=\frac{1+u-v+\sqrt{-4u+(1+u-v)^2}}{2}$,
           $\bar{z}=\frac{1+u-v-\sqrt{-4u+(1+u-v)^2}}{2}$
       } \\[6pt]
       &  $\bar{z}\to 0, z\to 0$ & {\footnotesize 
           $z=\frac{1+u-v-\sqrt{-4u+(1+u-v)^2}}{2}$,
           $\bar{z}=\frac{1+u-v+\sqrt{-4u+(1+u-v)^2}}{2}$
       } \\
       \hline\hline
    \end{tabular}
    \caption{The map between limit of $u,v$ and limit of $z,\bar{z}$. $\bar{z}\to 0, z\to 0$ means first taking $\bar{z}\to 0$, and then $z\to 0$. Here $u=z\bar{z}$ and $v=(1-z)(1-\bar{z})$.}
    \label{tab:map}
\end{table}
The branch choices for different limit cases (which may be used when the external legs are permuted and $u,v$ have different relations with $z,\bar{z}$) should also be chosen carefully to be consistent, we list our choice for all the cases in Table~\ref{tab:limchoice}.
\begin{table}[htbp]
    \centering
    \begin{tabular}{c|c|c|c}
    \hline\hline
       \thead{permutation \\ of external legs}  & \thead{transformation \\ of $u,v$} & \thead{expressions \\ of $z,\bar{z}$}
         & limits taken\\
         \hline
         1234 & $(u,v)$ & {\footnotesize 
           \thead{$z=\frac{1+u-v+\sqrt{-4u+(1+u-v)^2}}{2}$ \\
           $\bar{z}=\frac{1+u-v-\sqrt{-4u+(1+u-v)^2}}{2}$}
       } & $u\to 0, Y\to 0^{+}$ \\
       \hline 
       2134 & $(\frac{u}{v},\frac{1}{v})$ & {\footnotesize 
           \thead{$z=\frac{-1+u+v+\sqrt{-4u+(1+u-v)^2}}{2v}$ \\
           $\bar{z}=\frac{-1+u+v-\sqrt{-4u+(1+u-v)^2}}{2v}$}
       } & $u\to 0, Y\to 0^{-}$ \\
       \hline
       1324 & $(\frac{1}{u},\frac{v}{u})$ & {\footnotesize 
           \thead{$z=\frac{1+u-v-\sqrt{-4u+(1+u-v)^2}}{2u}$ \\
           $\bar{z}=\frac{1+u-v+\sqrt{-4u+(1+u-v)^2}}{2u}$}
       } & $u\to 0, Y\to 0^{+}$ \\
       \hline
       2314 & $(\frac{v}{u},\frac{1}{u})$ & {\footnotesize 
           \thead{$z=\frac{-1+u+v-\sqrt{-4u+(1+u-v)^2}}{2u}$ \\
           $\bar{z}=\frac{-1+u+v+\sqrt{-4u+(1+u-v)^2}}{2u}$}
       } & $u\to 0, Y\to 0^{-}$ \\
       \hline
       3124 & $(\frac{1}{v},\frac{u}{v})$ & {\footnotesize 
           \thead{$z=\frac{1-u+v+\sqrt{-4u+(1+u-v)^2}}{2v}$ \\
           $\bar{z}=\frac{1-u+v-\sqrt{-4u+(1+u-v)^2}}{2v}$}
       }  & $u\to 0, Y\to 0^{+}$\\
       \hline
       3214 & $(v,u)$ & {\footnotesize 
           \thead{$z=\frac{1-u+v+\sqrt{-4u+(1+u-v)^2}}{2}$ \\
           $\bar{z}=\frac{1-u+v-\sqrt{-4u+(1+u-v)^2}}{2}$}
       } & $u\to 0, Y\to 0^{-}$\\
       \hline\hline
    \end{tabular}
    \caption{The limits taken for different permutations of external legs. The first row is the default choice of external legs, which is named in order $1,2,3,4$. Our above choice is consistent in the sense that it guarantees that $\bar{z}$ approaches $0,1,\infty$ faster than $z$, and $z$ approaches $0,1$ from the right hand, approaches $\infty$ from the positive axis. }
    \label{tab:limchoice}
\end{table}

Second, though we have given our choices for the limit in $z,\bar{z}$, the final boundary result does not depend on this. They are expressions of $u$ and $v$ and more importantly, finite, which means that they have no poles in $z$ and $\bar{z}$ at all. The poles from leading singularity like $1/(z-\bar{z})$ and $1/(z-\bar{z})^2$ are spurious such that the conformal integrals are well defined on real axis. Thus whether $z$ or $\bar{z}$ approaching faster to the limit value does not matter for final results. This can also be seen directly from the boundaries calculated in the last section. The intriguing thing is that this simple property can provide us with more constraints on the possible analytic form of the integrals!

Let us explain in more detail with a simple example. For expression with double pole 
\begin{equation}
    \frac{1}{(z-\bar{z})^2}=\frac{1}{-4 u + u^2 + 2 u Y + Y^2},
\end{equation}
whether $z$ or $\bar{z}$ approaches 0 faster certainly matters. We can perform series expansion to both sides first with the assumption $0<\bar{z}<z$, then
\begin{equation}\label{eq:ztou}
    \frac{1}{z^2}+2\frac{\bar{z}}{z^3}+\ldots = \frac{1}{Y^2}-\left(\frac{2}{Y^3}-\frac{4}{Y^4}\right)u+\ldots
\end{equation}
The relation between $z,\bar{z}$ and $u,v$ is just the same as the first of the four rows in the last column of Table~\ref{tab:map}, the power of $\bar{z}$ directly corresponds to the power of $u$ as we indicated in \eqref{eq:ztou}.
If we further expand the left side into expressions of $u$ and $Y$ with the consistent choice we have made earlier, then it will agree with the right hand side perfectly. However, if we have used the ``inconsistent'' relation, for example, the second of the four rows in the last column from Table~\ref{tab:map} which indicates $z$ approaches 0 first, we will get a totally different result that seems to be nonsense 
\begin{equation}
    -7-\frac{1}{Y^2}+8Y+\frac{2Y^4}{u^3}+\frac{-7Y^2+8Y^3}{u^2}+\frac{2-14Y+12Y^2}{u}+u\left(2-\frac{4}{Y^4}+\frac{2}{Y^3}\right)+\ldots
\end{equation}
So for expressions which contain poles in $z,\bar{z}$, we indeed need to take care of the limit choice. However, for an expression which is free from such a pole, for instance, the following one
\begin{equation}
    \frac{z^2}{(z-\bar{z})^2}-\frac{2 z \bar{z}}{(z-\bar{z})^2}+\frac{\bar{z}^2}{(z-\bar{z})^2}=1.
\end{equation}
We have split it into three parts, each part has the double pole. Again we first expand with the assumption $0<\bar{z}<z$:
\begin{equation}\label{eq:zexpand}
    \begin{aligned}
        \frac{z^2}{(z-\bar{z})^2}&= 1+2\frac{\bar{z}}{z}+3\frac{\bar{z}^2}{z^2}+\cdots,\,
        \frac{-2z\bar{z}}{(z-\bar{z})^2}= -2\frac{\bar{z}}{z}-4\frac{\bar{z}^2}{z^2}+\cdots,\,
        \frac{\bar{z}^2}{(z-\bar{z})^2}=\frac{\bar{z}^2}{z^2} +\cdots
    \end{aligned}
\end{equation}
then use the consistent choice for $z,\bar{z}$ and expand to $\mathcal{O}(u^0,Y^3)$
\begin{equation}\label{eq:consistent}
    \begin{aligned}
        \frac{z^2}{(z-\bar{z})^2}&= 1+\mathcal{O}(u^1), \,
        \frac{-2z\bar{z}}{(z-\bar{z})^2}= \mathcal{O}(u^1),\,
        \frac{\bar{z}^2}{(z-\bar{z})^2}= \mathcal{O}(u^2)
    \end{aligned}
\end{equation}
Meanwhile, we use the ``inconsistent'' choice for $z,\bar{z}$, expand to $\mathcal{O}(u^0,Y^3)$ as well
\begin{equation}\label{eq:inconsistent}
    \begin{aligned}
        \frac{z^2}{(z-\bar{z})^2}&\to \frac{\mathcal{O}(Y^4)}{u^2}+\frac{-10Y^2+12Y^3+\mathcal{O}(Y^4)}{u}+\left(3-20Y+18Y^2+\mathcal{O}(Y^4)\right)+\mathcal{O}(u^1), \, \\
        \frac{-2z\bar{z}}{(z-\bar{z})^2}&\to \frac{\mathcal{O}(Y^4)}{u^2}+\frac{14Y^2-16Y^3+\mathcal{O}(Y^4)}{u}+\left(-4+28Y-24Y^2+\mathcal{O}(Y^4)\right)+\mathcal{O}(u^1),\, \\
        \frac{\bar{z}^2}{(z-\bar{z})^2}&\to \frac{\mathcal{O}(Y^4)}{u^2}+\frac{-4Y^2+4Y^3+\mathcal{O}(Y^4)}{u}+\left(2-8Y+6Y^2+\mathcal{O}(Y^4)\right)+\mathcal{O}(u^1).
    \end{aligned}
\end{equation}
Here we use $\to$ instead of $=$ because we use expression in \eqref{eq:zexpand} which truncates to $\bar{z}^2$. We can see that no matter we choose the consistent or ``inconsistent'' expressions, both add up to the correct exact result up to the expansion order. 

So the ``inconsistent'' choice can also be used if the results are free of poles of $z,\bar{z}$. The next question is why the ``inconsistent'' choice can give us more constraints? The reason is that the boundaries we give to match our ansatz are truncated to $\mathcal{O}(u^0)$. For the above example, suppose we want to find an expression which expands to the boundary as follows
\begin{equation}
    \left.c_1\frac{z^2}{(z-\bar{z})^2}-c_2\frac{2 z \bar{z}}{(z-\bar{z})^2}+c_3\frac{\bar{z}^2}{(z-\bar{z})^2}\right|_{u\to0,Y\to 0^{+}}=1+\mathcal{O}(u^1).
\end{equation}
If we use the consistent expressions for $z,\bar{z}$ which gives \eqref{eq:consistent}, then we can only get $c_1=1$. It can fix all of them only when we expand to $\mathcal{O}(u^2)$. However, the ``inconsistent'' choice, that is, \eqref{eq:inconsistent} will immediately give us a set of constraints
\begin{equation}
    \begin{aligned}
        3c_1-4c_2+2c_3=&1, \, -20c_1+28c_2-8c_3=0, 18c_1-24c_2+6c_3=0, \\
        -10c_1+14c_2-4c_3=&0,\, 12c_1-16c_2+4c_3=0
    \end{aligned}
\end{equation}
Then we fix all three coefficients $c_1=c_2=c_3=1$. Both of them use the same boundary condition. This illustrates the power of analytic properties of finite conformal integrals. We can use both the consistent and ``inconsistent'' expressions for the series expansion of our ansatz. And we indeed find in some examples (see Sec.~\ref{sec:fourloopconformal}) that only when the constraints from ``inconsistent'' choice are added can we fix all the coefficients from boundary conditions alone. Since the ``inconsistent'' choice exchanges the expressions of $z$ and $\bar{z}$, we will also call the constraints solved from them mirror constraints for simplicity.


\section{Bootstrap for conformal integrals}\label{sec:conformal}
In this section we try to apply the bootstrap method set up above to general four-point conformal integrals that may appear in $\mathcal{N}=4$ super-Yang-Mills at three and four loops. We first define the scope of what we mean by general conformal integrals. These conformal integrals mainly appear in the perturbative calculation of four-point correlation functions of half-BPS operators $\langle O_{p}(x_1)O_{q}(x_2)O_{r}(x_3)O_{s}(x_4)\rangle$, we refer to~\cite{Heslop:2022xgp} for a comprehensive review of this physical object and its relation to scattering amplitudes. Due to the hidden permutation symmetry of four-point correlation functions~\cite{Eden:2011we}, their integrands are more conveniently described by so-called $f$-graphs which correspond to
\begin{equation}\label{eq:fint}
    f^{(l)}(x_1,\ldots,x_{l+4})=\frac{P^{(l)}(x_1,\ldots,x_{l+4})}{\prod_{1\le i<j\le l+4}x_{ij}^{2}}
\end{equation}
where $P^{(l)}$ is a polynomial with permutation symmetry of all $l+4$ vertices and is of degree $l-1$ of each vertex. Such graphs are constrained by the following several conditions. First, the conformal symmetry of the theory requires the degree of each vertex to be 4. Second, the simple pole assumption\footnote{This assumption originates from the analysis of OPE limit~\cite{Eden:2011we}.} requires that there are no double or higher poles like $1/(x_{ij}^{2})^{2}$ in the integrand. Then the hidden permutation symmetry allows one to use a single graph to represent all the permutations of the labels of vertices. Graphs with such properties are few in lower loops and not all of them will contribute to the correlation function. If we consider the planar limit\footnote{The planar limit here is $N_{c}\to \infty$ where $N_c$ corresponds to the $SU(N_c)$ gauge group. In planar limit, the denominator factors of those $f$-graphs which contribute to correlators must form a plane graph. We will call those graphs without plane embeddings the nonplanar $f$-graphs.}, such graphs and how they contribute to correlators have been determined up to 12 loops using graphical rules~\cite{Bourjaily:2025iad,Bourjaily:2011hi,Bourjaily:2016evz,He:2024cej}.

However, hereafter we do not restrict ourselves to planar limit, that is, we will also include all nonplanar $f$-graphs. Such $f$-graphs will contribute to correlation function starting from four loops~\cite{Eden:2012tu}, and how they contribute to the integrand is well studied at four loop~\cite{Fleury:2019ydf} and five loop~\cite{Bargheer:2025hvd}. For one and two loops, there are no such nonplanar $f$-graphs. For three loops, their contribution is cancelled by Gram identity. Nevertheless, these nonplanar graphs still contribute to other kinds of correlators like the mixed heavy-light four-point correlator $\langle GGOO\rangle$ involving giant graviton operators starting at three loops~\cite{He:2026ios}. For such correlators, nonplanar graphs contribute even in the large-$N_c$ limit. In the following, we will generate all the possible $f$-graphs satisfying conformality and locality for three-loop and four loops\footnote{It is relatively easy to generate the numerators first, that is, $P^{(l)}$ in \eqref{eq:fint} using python package \texttt{nauty} and command like \texttt{geng\ \textless{}l+4\textgreater{}\ |\ multig\ -m\textless{}l-1\textgreater{}\ -A\ -r\textless{}l-1\textgreater{}\ \textgreater{}\ pgraphs\_l\textless{}l\textgreater{}.txt} where $l$ is the loop order. They can also be described by a set of graphs called $P$-graphs in~\cite{Eden:2012tu}.}. Once they are generated, we then take four vertices of these graphs to be external (they will always be named $x_{1},x_2,x_3,x_4$) and find all the inequivalent integrands generated. Here `inequivalent' means two graphs can not change to each other by permutation of labels of four external vertices \textit{or} the permutation of labels of internal vertices (they are loop vertices and will be integrated out).

\subsection{Three loop conformal integrals}\label{sec:threeloopconformal}
At three loop, there are totally 4 $f$-graphs, one is planar and the rest are nonplanar. We choose four points as external vertices and get 15 inequivalent conformal integrals, which agrees with~\cite{He:2026ios}. They are depicted in Appendix~\ref{app:threeloopint}. However, there are identities between conformal integrals which further reduce independent functions. 

The first kind is the magic identity~\cite{Drummond:2006rz}. It will generate two identities
\begin{equation}
    \mathcal{I}^{(3)}_{2}\Bigl(\frac{z}{z-1}\Bigr)=-\mathcal{I}^{(3)}_{3}(z), \quad \mathcal{I}^{(3)}_{9}(1-z)=-\mathcal{I}^{(3)}_{11}(z).
\end{equation}
The difference of arguments originates from the different ordering of external vertices in the definition of Fig.~\ref{fig:threeloopbasis}. When we perform the permutation of external vertices, it induces a M\"obius transformation of $z$ and $\bar{z}$. Some integrals have additional symmetries under the permutation of external vertices, we also list all of them here for later convenience.
For $i\in\{1,5,10,11,13\}$,
\begin{equation}\label{eq:symeq1}
    \mathcal{I}^{(3)}_{i}(z)=\mathcal{I}^{(3)}_{i}\Bigl(\frac{z}{z-1}\Bigr),\quad
    \mathcal{I}^{(3)}_{i}(1-z)=\mathcal{I}^{(3)}_{i}\Bigl(1-\frac{1}{z}\Bigr),\quad
    \mathcal{I}^{(3)}_{i}\Bigl(\frac{1}{z}\Bigr)=\mathcal{I}^{(3)}_{i}\Bigl(\frac{1}{1-z}\Bigr).
\end{equation}
For $i\in\{4,9,14\}$,
\begin{equation}\label{eq:symeq2}
    \mathcal{I}^{(3)}_{i}(z)=\mathcal{I}^{(3)}_{i}\Bigl(\frac{1}{z}\Bigr),\quad
    \mathcal{I}^{(3)}_{i}(1-z)=\mathcal{I}^{(3)}_{i}\Bigl(\frac{1}{1-z}\Bigr),\quad
    \mathcal{I}^{(3)}_{i}\Bigl(1-\frac{1}{z}\Bigr)=\mathcal{I}^{(3)}_{i}\Bigl(\frac{z}{z-1}\Bigr).
\end{equation}
For $i\in\{7,15\}$, they are totally symmetric
\begin{equation}\label{eq:symeq3}
    \mathcal{I}^{(3)}_{i}(z)=\mathcal{I}^{(3)}_{i}(1-z)=\mathcal{I}^{(3)}_{i}\Bigl(1-\frac{1}{z}\Bigr)=\mathcal{I}^{(3)}_{i}\Bigl(\frac{1}{z}\Bigr)=\mathcal{I}^{(3)}_{i}\Bigl(\frac{1}{1-z}\Bigr)=\mathcal{I}^{(3)}_{i}\Bigl(\frac{z}{z-1}\Bigr).
\end{equation}
For $i\in\{6,12\}$,
\begin{equation}\label{eq:symeq4}
    \begin{aligned}
        \frac{1}{(1-z)(1-\bar{z})}\mathcal{I}^{(3)}_{i}(z) &= \mathcal{I}^{(3)}_{i}\Bigl(\frac{z}{z-1}\Bigr),\, \mathcal{I}^{(3)}_{i}(1-z) &= z\bar{z}\,\mathcal{I}^{(3)}_{i}\Bigl(1-\frac{1}{z}\Bigr),\\
        \mathcal{I}^{(3)}_{i}\Bigl(\frac{1}{1-z}\Bigr) &= \frac{z\bar{z}}{(1-z)(1-\bar{z})}\,\mathcal{I}^{(3)}_{i}\Bigl(\frac{1}{z}\Bigr).
    \end{aligned}
\end{equation}
At last, we have
\begin{equation}\label{eq:symeq5}
    \begin{aligned}
        \mathcal{I}^{(3)}_{8}(z) &= \mathcal{I}^{(3)}_{8}\Bigl(\frac{z}{z-1}\Bigr)=\frac{1}{z\bar{z}}\,\mathcal{I}^{(3)}_{8}\Bigl(\frac{1}{z}\Bigr)=\frac{1}{z\bar{z}}\,\mathcal{I}^{(3)}_{8}\Bigl(\frac{1}{1-z}\Bigr)\\
        &=\frac{(1-z)(1-\bar{z})}{z\bar{z}}\,\mathcal{I}^{(3)}_{8}(1-z)=\frac{(1-z)(1-\bar{z})}{z\bar{z}}\,\mathcal{I}^{(3)}_{8}\Bigl(1-\frac{1}{z}\Bigr).
    \end{aligned}
\end{equation}
Actually, $\mathcal{I}^{(3)}_{6}$, $\mathcal{I}^{(3)}_{7}$ and $\mathcal{I}^{(3)}_{8}$ are reducible integrals which are products of lower-loop integrals. The second kind of identity is the Gram identity. At three loops, there are totally 7 points, thus we can construct our first Gram identity in the embedding space (which is six dimensional). It is what we have presented in \eqref{eq:gram}. We can take this Gram as the numerator $P^{(3)}$ in \eqref{eq:fint}, it has the correct degree 2 for all the vertices. By dividing with the denominator which is a $7$-point complete graph and integrate out internal vertices, we turn Gram identity into an integral identity. 
Applying the symmetry identities given in \eqref{eq:symeq1}--\eqref{eq:symeq5}, we arrive at the following compact identity for $\mathcal{I}^{(3)}_{15}(z)$,
\begin{equation}\label{eq:gramid}
    \begin{aligned}
        \mathcal{I}^{(3)}_{15}(z)&=P_3\bigl(\mathcal{I}^{(3)}_{1}\bigr) - 3\,S_6\bigl(\mathcal{I}^{(3)}_{2}\bigr) + 2\,Q_3\bigl(\mathcal{I}^{(3)}_{4}\bigr)+ P_{3}\left(\frac{-2+z+\bar{z}+z\bar{z}}{z\bar{z}}\,\mathcal{I}^{(3)}_{5}(z)\right) \\
        &-P_3\left((2-z-\bar{z})\,\mathcal{I}^{(3)}_{6}\Bigl(\frac{z}{z-1}\Bigr)\right) + 2\,\mathcal{I}^{(3)}_{7} - \frac{(z-\bar{z})^{2}}{12(1-z)(1-\bar{z})}\,\mathcal{I}^{(3)}_{8} \\
        &+ 2\,P_3\bigl(\mathcal{I}^{(3)}_{10}\bigr) + 3\,P_3\bigl(\mathcal{I}^{(3)}_{11}\bigr) +\frac{1}{2} P_{3}\left((2-z-\bar{z})\,\mathcal{I}^{(3)}_{12}\Bigl(\frac{z}{z-1}\Bigr)\right) - P_3\bigl(\mathcal{I}^{(3)}_{13}\bigr) - Q_3\bigl(\mathcal{I}^{(3)}_{14}\bigr).
    \end{aligned}
\end{equation}
where $\mathcal{I}^{(3)}_i$ is a shorthand for $\mathcal{I}^{(3)}_i(z)$ when the argument is trivial and we have also defined the following compact notation for sums over symmetry orbits:
\begin{equation}\label{eq:PQdef}
    \begin{aligned}
        P_3(f(z)) &\equiv f\Bigl(1-\frac{1}{z}\Bigr) + f\Bigl(\frac{1}{z}\Bigr) + f(z),\\
        Q_3(f(z)) &\equiv f\Bigl(\frac{1}{1-z}\Bigr) + f\Bigl(\frac{z}{z-1}\Bigr) + f(z),\\
        S_6(f(z)) &\equiv f\Bigl(1-\frac{1}{z}\Bigr) + f\Bigl(\frac{1}{1-z}\Bigr) + f(1-z) + f\Bigl(\frac{1}{z}\Bigr) + f(z) + f\Bigl(\frac{z}{z-1}\Bigr).
    \end{aligned}
\end{equation}
Though $\mathcal{I}^{(3)}_{15}(z)$ can not be obtained from \texttt{HyperlogProcedures} directly, all the integrals on the right hand side can be evaluated by this package. Therefore the analytic expression for $\mathcal{I}^{(3)}_{15}(z)$ can be obtained finally. We give its expression in the ancillary file. Then at three loop, we have already obtained analytic results for all the conformal integrals in the scope. 

Since three-loop results are all known, we can use them as benchmark to test the above workflow of bootstrap. And it indeed works for different leading singularities and we find that the boundary condition is sufficient up to order $\mathcal{O}(u^0,Y^3)$ and the consistent choice in Tab.~\ref{tab:limchoice} for the series expansion works fine. We provide three examples for $\mathcal{I}^{(3)}_{5}$, $\mathcal{I}^{(3)}_{8}$ (cubic of boxes) and $\mathcal{I}^{(3)}_{12}$ (hard integrals) in the repository. Then we go to the four-loop cases.

\subsection{Four loop conformal integrals}\label{sec:fourloopconformal}
At four loop, there are totally 32 $f$-graphs, only three of them are planar and the rest are nonplanar. All 3 planar $f$-graphs will contribute to the four-point correlators in the planar limit and 21 of them will contribute to the order $1/N_{c}^{2}$ (genus-1 corrections)~\cite{Fleury:2019ydf}. If we restrict ourselves to planar $f$-graphs, 32 independent integrands (in the sense that they can not be permuted to each other) will be generated. However, for our purpose, those integrands which are proportional to each other up to a constant factor of $u,v$ are also taken as dependent. This is what we mean by `independent' hereafter. So we actually have 31 independent integrands. If we take all the $32$ $f$-graphs into account, there are totally 412 such four-loop conformal integrals. Here we take all these 412 conformal integrals into consideration. Since there are too many of them, all their integrands are provided in the ancillary files. We also provide the indices for the 32 integrals that contribute in planar limit in our larger basis in case some reader may be interested in the planar contribution. They are 
\begin{equation}\label{eq:planarlist}
    \begin{aligned}
        &\mathcal{I}^{(4)}_{3},\; \mathcal{I}^{(4)}_{5},\; \mathcal{I}^{(4)}_{21},\; \frac{(1-z)(1-\bar{z})}{z\bar{z}}\mathcal{I}^{(4)}_{27},\; \mathcal{I}^{(4)}_{37},\; (1-z)(1-\bar{z})\mathcal{I}^{(4)}_{38},\; \mathcal{I}^{(4)}_{39},\; \frac{(1-z)(1-\bar{z})}{z\bar{z}}\mathcal{I}^{(4)}_{49},\\
        &\frac{(1-z)^2(1-\bar{z})^2}{z\bar{z}}\mathcal{I}^{(4)}_{49},\; \frac{z\bar{z}}{(1-z)(1-\bar{z})}\mathcal{I}^{(4)}_{126},\; \frac{z\bar{z}}{(1-z)(1-\bar{z})}\mathcal{I}^{(4)}_{133},\; (1-z)(1-\bar{z})\mathcal{I}^{(4)}_{140},\; \\
        &z\bar{z}\,\mathcal{I}^{(4)}_{144},\; \mathcal{I}^{(4)}_{160},\; \mathcal{I}^{(4)}_{162},\; \mathcal{I}^{(4)}_{163},\;\mathcal{I}^{(4)}_{164},\; \mathcal{I}^{(4)}_{167},\; \mathcal{I}^{(4)}_{168},\; z\bar{z}\,\mathcal{I}^{(4)}_{169},\; \mathcal{I}^{(4)}_{173},\; \mathcal{I}^{(4)}_{174},\; \mathcal{I}^{(4)}_{175},\; \mathcal{I}^{(4)}_{176},\\
        &\mathcal{I}^{(4)}_{181},\; \mathcal{I}^{(4)}_{187},\; \mathcal{I}^{(4)}_{208},\; \mathcal{I}^{(4)}_{212},\; \frac{\mathcal{I}^{(4)}_{214}}{(1-z)(1-\bar{z})},\; \frac{1}{z\bar{z}}\mathcal{I}^{(4)}_{233},\; \mathcal{I}^{(4)}_{246},\; \frac{z\bar{z}}{(1-z)(1-\bar{z})}\mathcal{I}^{(4)}_{260}.
    \end{aligned}
\end{equation}
where $\mathcal{I}_{\ast}^{(4)}$ is a shorthand for $\mathcal{I}_{\ast}^{(4)}(z)$. The basis we choose may differ from the integrand that contributes by a constant factor, which is the reason for the appearance of factors like $z\bar{z}$ before $\mathcal{I}^{(4)}_{144}$. The integral $\mathcal{I}^{(4)}_{168}$ in Sec.\ref{sec:ls} that contains elliptic cut is also one of them. Then we consider the integral identities as in the three-loop case.

First, we find only 39 magic relations which relate different pair of integrals among the 412 basis we consider\footnote{Here we do not consider the new magic identity which is more magic and thus rare in four loops, see~\cite{Caron-Huot:2021usw,He:2025vqt} for such generalized identities which are obtained by matching with integrability data.}. Second, we obtain 686 symmetries identities for $213$ integrals. That is, about one half of the integrals in basis have discrete symmetries by permuting external legs. At last, we consider the Gram identities for four-loop conformal integrals. Such Gram identities have shown their power in reducing the periods of nonplanar $f$-graphs at four loop~\cite{Zhang:2024ypu}. Since for four loops there are 8 different vertices, we can actually generate 36 different Gram identities $|x_{ij}^2|$, where $i,j \in\{1,2,3,4,5,6,7\}$ or  $i \in\{1,2,3,4,5,6,7\}$ and $j \in\{1,2,3,4,5,6,8\}$, including all the permutations of indices\footnote{Here we are not dealing with identities between $f$-graphs, which are permutation invariant themselves. For conformal integrals, those permutations are different, so the total number is $C^{1}_{8}+C^{2}_8=36$.}. These Gram identities will be put in the numerator, and the most general denominator satisfying locality is the complete graph $\prod_{1\le i<j\le l+4}x_{ij}^{2}$.
To make this expression conformally invariant, we still need an additional monomial which multiplies the Gram identity in the numerator. It must be of degree $-1$ for vertices appearing twice in the list of $i,j$, of degree $-2$ for vertices appearing once and of degree $-3$ for those not appearing in Gram at all. Due to these strong constraints from conformal weight, we can generate all possible cases and thus all the integral identities related to Gram identities. There are 710 of them. However, we have observed that after substituting the magic identities and symmetry conditions for integrals, the Gram identities have reduced a lot and only few of them are independent. If we take $\mathcal{I}^{(4)}_{i}(z)$ and its permutation of external vertices as independent variables (except for those related by symmetry identities), then there are only 69 variables (related to 26 different conformal integrals) can be solved by Gram identities at last. Thus the power of Gram identities become weaker with the increasing of loop orders. This is expected since the number of independent integrals increases exponentially.

However, we find that 164 of the 412 integrals can be directly computed by the package \texttt{HyperlogProcedures}. This proves the powerfulness of the graphical function method for conformal integrals with numerators~\cite{Schnetz:2025opm}. Nevertheless, in the following, we will focus on several examples that can not be calculated by this package. Their results can only be obtained through bootstrap to our knowledge. We mainly use them to show several analytic properties presented in Sec.~\ref{sec:ls}.

\subsubsection{Solving using mirror constraints}
The first example is the integral $\mathcal{I}^{(4)}_{173}$ which has been bootstrapped in~\cite{He:2025vqt}\footnote{This integral is named differently as $\mathcal{I}^{f}$ in~\cite{He:2025vqt}.}. Here we give more details of the bootstrap used. $\mathcal{I}^{(4)}_{173}$ has two leading singularities, $1/(z-\bar{z})$ and $(u-v-1)/(z-\bar{z})^2$. Then its ansatz can be written as
\begin{equation}\label{eq:ansatzI173}
    \mathcal{I}^{(4)}_{173}=\sum_{i}c_i\frac{\mathcal{L}^{\text{odd}}_{\text{ans},i}}{z-\bar{z}}+\sum_{j}d_j\frac{-2+z+\bar{z}}{(z-\bar{z})^2}\mathcal{L}^{\text{even}}_{\text{ans},j}
\end{equation}
where $\mathcal{L}^{\text{odd}}_{\text{ans},i}$ and $\mathcal{L}^{\text{even}}_{\text{ans},j}$ are parity-odd and parity-even (under $z\leftrightarrow\bar{z}$) weight-8 svHPLs. This function space can be inferred either by the empirical rule in App.~\ref{app:reducerule} or the new magic identity~\cite{Caron-Huot:2021usw,He:2025vqt} satisfied by this integral. There are 135 parity-odd basis and 169 parity-even basis. 

If we use the consistent choice for expressions of $z,\bar{z}$ in Table~\ref{tab:limchoice} and match with the boundary data computed, it can determine 221 of the 304 parameters, leaving 83 undetermined. If we further use the new magic identity~\cite{Caron-Huot:2021usw,He:2025vqt} for this integral, we can fix all of them. The new magic identity plays the role that it splits the contribution from two different leading singularities of $\mathcal{I}^{(4)}_{173}$.
However, here we want to show that how the ``inconsistent'' choice of expressions of $z,\bar{z}$ in Sec.~\ref{sec:series} can give us more constraints to directly fix all the coefficients by boundary conditions alone. That is, the boundary data already contains all the necessary information, the new magic identity is not necessary for the bootstrap. Since the ``inconsistent'' choice exchanges the algebraic expressions of $z,\bar{z}$, we will also call the constraints solved from them mirror constraints.

We take the series expansion around $z=0$ as an example. We first expand the ansatz with the assumption that $0<\bar{z}<z$. Then the denominators are expanded to series of $\frac{\bar{z}}{z}$. The whole ansatz turns into a series expansion with the negative (and positive) orders of $z$ and positive orders of $\bar{z}$. Then we use the following expressions of $z$ and $\bar{z}$ which is apparently against the first assumption $0<\bar{z}<z$,
\begin{equation}
    z=\frac{1+u-v-\sqrt{-4u+(1+u-v)^2}}{2}, \quad \bar{z}=\frac{1+u-v+\sqrt{-4u+(1+u-v)^2}}{2}
\end{equation}
with the same limit $u\to 0,Y\to 0^{+}$ as the consistent choice. However, just as our simple example in Sec.~\ref{sec:series} shows, this ``illegal'' usage is actually permitted because the final result of the conformal integrals have no pole or branch cut on the real line $z=\bar{z}$. The mirror matching with the same boundary conditions gives us 121 constraints, which is greater than 89 using the consistent choice. Similarly, for other five different boundaries and corresponding limits, we can do the same thing and at last all 304 parameters are fixed and agree with those obtained by imposing new magic identity. The numerical value of this integral has been checked already in~\cite{He:2025vqt}, here we check it again using \texttt{trillo}~\cite{trillo}.

\subsubsection{Solving by splitting leading singularities}
The second example is $\mathcal{I}^{(4)}_{176}$ whose leading singularity has been studied in Sec.~\ref{sec:splitls}. We have split the original integrand into three parts, each of which can be bootstrapped in an easier way since there is only single leading singularity. The construction of its ansatz has also been discussed in Sec.~\ref{sec:ansatz}. Thus we are ready to bootstrap its results and $\mathcal{I}^{a,b,c}_{176}$\footnote{We apologize here for different notations for the same object in the data provided as ancillary files. In our 412 basis, $\mathcal{I}^{b}_{176}$ is actually $\frac{1}{z\bar{z}}\mathcal{I}^{(4)}_{117}(\frac{1}{z})$ and $\mathcal{I}^{c}_{176}$ is $\mathcal{I}^{(4)}_{120}(z)$. They may differ from the definition in our basis by the permutation of external vertices or some constant factor.} are successfully fixed. $\mathcal{I}^{(4)}_{176}=\mathcal{I}^{a}_{176}+\mathcal{I}^{b}_{176}$ and $\mathcal{I}^{b}_{176}$ both contain the new letter $1/\bar{z}$ in the last two entries, while $\mathcal{I}^{c}_{176}$ does not contain new letter and is totally composed of svHPLs. 

\begin{figure}
    \centering
    \includegraphics[width=0.3\textwidth]{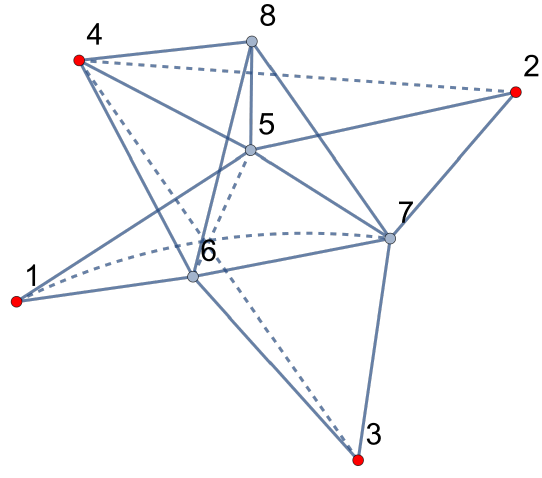}
    \caption{Another four-loop conformal integral which is of easy type, because it has the same leading singularities as three-loop easy integral.}\label{fig:I4L181}
\end{figure}

Now we can use the same method to solve a similar integral $\mathcal{I}^{(4)}_{181}$,
\begin{equation}
    \mathcal{I}^{(4)}_{181}=\frac{x_{17}^2 x_{24}^2 x_{34}^2 x_{56}^2}{x_{15}^2 x_{16}^2 x_{25}^2 x_{27}^2 x_{36}^2 x_{37}^2 x_{45}^2 x_{46}^2 x_{48}^2 x_{57}^2 x_{58}^2 x_{67}^2 x_{68}^2 x_{78}^2}.
\end{equation}
which is depicted in Fig.~\ref{fig:I4L181}.
We note that both $\mathcal{I}^{(4)}_{176}$ and $\mathcal{I}^{(4)}_{181}$ are of easy type at four loop, because they have the same leading singularities as three-loop easy integral. More interestingly, they can actually be reduced to three-loop easy integral using the empirical rule presented in App.~\ref{app:reducerule}, this is another reason we classify them as easy type. The leading singularity analysis traces one of its leading singularities $1/(z-\bar{z})$ back to the following two cuts of $\mathcal{I}^{(4)}_{181}$:
\begin{equation}
    \begin{array}{cc}
        \text{cut 1: }&x_{15}^{2}, \, x_{25}^{2}, \, x_{45}^{2}, x_{47}^{2}x_{56}^{2}-x_{46}^{2}x_{57}^{2}, \\
        \text{cut 2: }&x_{15}^{2}, \, x_{25}^{2}, \, x_{57}^{2}, x_{47}^{2}x_{56}^{2}-x_{45}^{2}x_{67}^{2}.
    \end{array}
\end{equation}
where the last composite expressions come from Jacobian factors. Then we can construct such a numerator $x_{47}^{2}x_{56}^{2}-x_{46}^{2}x_{57}^2-x_{45}^{2}x_{67}^{2}$. It has the nice property that, under first cut, it cancels $x_{47}^{2}x_{56}^{2}-x_{46}^{2}x_{57}^{2}$ in the denominator and under the second cut, it cancels $x_{47}^{2}x_{56}^{2}-x_{45}^{2}x_{67}^{2}$. Thus this is the correct numerator to eliminate leading singularity $1/(z-\bar{z})$ from $\mathcal{I}^{4}_{181}$. Then we construct the first integral
\begin{equation}
    \mathcal{I}^{a}_{181}=\frac{x_{13}^{2}x_{24}^{2}(x_{47}^{2}x_{56}^{2}-x_{46}^{2}x_{57}^2-x_{45}^{2}x_{67}^{2})}{x_{15}^2 x_{16}^2 x_{25}^2 x_{27}^2 x_{36}^2 x_{37}^2 x_{45}^2 x_{46}^2 x_{48}^2 x_{57}^2 x_{58}^2 x_{67}^2 x_{68}^2 x_{78}^2}
\end{equation}
which has a single leading singularity $1/(1-u)/(z-\bar{z})$. Then we subtract the part with leading singularity $1/(1-u)/(z-\bar{z})$ from the original integral to get another part with single leading singularity $1/(z-\bar{z})$. However, notice that $\mathcal{I}^{(4)}_{181}$ has the symmetry with exchange of $2\leftrightarrow 3$, thus we should subtract $\mathcal{I}^{a}_{181}$ and its image under $2\leftrightarrow 3$. This also agrees with the leading singularity analysis, since there are actually two leading singularities $u/(1-u)/(z-\bar{z}), 1/(1-u)/(z-\bar{z})$ symmetric to each other. Then we obtain the second integral
\begin{equation}\label{eq:I4L181b}
    \mathcal{I}^{b}_{181}=\frac{2x_{17}^2 x_{24}^2 x_{34}^2 x_{56}^2\!-\!x_{13}^{2}x_{24}^{2}(x_{47}^{2}x_{56}^{2}\!-\!x_{46}^{2}x_{57}^2\!-\!x_{45}^{2}x_{67}^{2})\!-\!x_{12}^{2}x_{34}^{2}(x_{47}^{2}x_{56}^{2}\!-\!x_{46}^{2}x_{57}^2\!-\!x_{45}^{2}x_{67}^{2})}{2x_{15}^2 x_{16}^2 x_{25}^2 x_{27}^2 x_{36}^2 x_{37}^2 x_{45}^2 x_{46}^2 x_{48}^2 x_{57}^2 x_{58}^2 x_{67}^2 x_{68}^2 x_{78}^2}.
\end{equation} 
with single leading singularity $1/(z-\bar{z})$. We can carry out the bootstrap procedure for both of them, then we can derive
\begin{equation}
    \mathcal{I}^{(4)}_{181}=\mathcal{I}^{b}_{181}+\frac{1+u}{2}\mathcal{I}^{a}_{181}.
\end{equation}
All the numerical values of the four-loop integrals we calculated as examples are checked by \texttt{trillo} up to $10^{-4}\sim 10^{-5}$ relative error. The reader can check the ancillary \texttt{Mathematica} notebook for more details about their analytic expressions and numerical check.

\section{Conclusions}
In this note, we set up a complete workflow for the bootstrap of three-loop and four-loop conformal integrals which are evaluated to svMPLs. These conformal integrals are also a subclass of graphical functions. We mainly exploit the analytic information from three aspects: first the leading singularities and the cut information in intermediate steps, second the six boundary conditions by taking two external points approaching to each other, and at last the nice property of finite conformal integrals that they have no pole or branch cuts on the real axis. It is interesting that boundary conditions have such strong constraints for finite conformal integrals with so limited expansion orders. We have also observed an empirical rule that has some interesting correspondence with boxing differential equation which may provide more information of letters in deeper depth. 
Finally, a package with some skill files which can be easily used by current AI models is provided. Detailed information of the package and how to use it are all summarized as markdown files in the git repository \faGithub\, \url{https://github.com/windfolgen/svbwalkthrough.git}. It sets up the calculation in bootstrap workflow and gives the leading singularity analysis of three-loop and four-loop conformal integrals (though in a rough way).

However, we also remind the reader that such bootstrap method is actually very limited for four-loop conformal integrals. The main reason is that most conformal integrals with a graphical representation are not rigid. They are not of uniform weight or may involve so many different leading singularities and contain so many different components in their expressions which greatly weakens the power of bootstrap method. This method is more suitable for integrals which have single or two leading singularities, no matter how complicated its integrand will be. $\mathcal{I}^{b}_{181}$ in \eqref{eq:I4L181b} is such an example. So it may work better for those conformal integrals that are specially designed to have good cut properties as in~\cite{He:2025rza}, though our initial aim is to study the graphical functions. Nevertheless, we still obtain some new results of four-loop conformal integrals. We find the expressions of some are simple, like $\mathcal{I}^{c}_{176}$, which may indicate that there is an easier way to calculate such integrals.
Another direction is that starting from four loops, we find many conformal integrals containing elliptic curves in the cut. Such integrals are the main obstacles to obtaining the full analytic results for the four-loop four-point correlators in $\mathcal{N}=4$ SYM. And it still remains unclear how single-valued elliptic polylogarithms~\cite{Schlotterer:2025qjv,Baune:2025ndr} are related to such conformal integrals or whether we can construct such a function space like svMPLs for these integrals.
\acknowledgments

The author thanks Shun-Qing Zhang for providing the numerical program \texttt{trillo} to check the numerical results. He also thanks Song He, Canxin Shi, Jiahao Liu and Qinglin Yang for useful discussions on related problems and helpful suggestions on the manuscript. Finally, the author thanks Song He for his support and encouragement in writing this note. This work is supported by National Natural Science Foundation of China under Grant No. 12225510.

\appendix
\section{From boxing differential equation to graph reduction}\label{app:boxing}
The boxing operator acting on the integrand takes the following form (we use the Euclidean metric here),
\begin{equation}
    \square_{3}\frac{1}{x_{35}^2}=\partial_{x_3^{\mu}}\partial_{x_{3\mu}}\frac{1}{x_{35}^2}=-\partial_{x_3^{\mu}}\frac{2x_{35}^{\mu}}{(x_{35}^2)^2}
\end{equation}
where $x_3$ is an external vertex and $x_5$ an internal vertex that is to be integrated out. $x_{35}$ is a shorthand for $x_3-x_5$. If the external vertex is in $d$ dimension, we will have
\begin{equation}\label{eq:intboxing}
    \square_{3}\frac{1}{x_{35}^2}=(8-2d)\frac{1}{(x_{35}^2)^2}
\end{equation}
which vanishes in $d=4$ dimension. However, since this is an integration of $x_5$, we also need to consider the singular point where $x_3=x_5$. Just like the Cauchy residue theorem, we take a sphere around $x_5$ with radius $r$ and $r\to 0$ to take this into account.
\begin{equation}
    -\lim_{r\to 0}\int_{V}\partial_{x_3^{\mu}}\frac{2x_{35}^{\mu}}{(x_{35}^2)^2}\mathrm{d}^{d}x_5=-\lim_{r\to 0}\int_{S}\frac{2r^d}{r^4}\mathrm{d}S=-\frac{4\pi^{d/2}}{\Gamma(\frac{d}{2})}\lim_{r\to 0}r^{d-4}
\end{equation}
So when $d=4$, such a contribution turns into a delta function
\begin{equation}\label{eq:boxing1}
    \square_{3}\frac{1}{x_{35}^2}=-4\pi^2\delta^{(4)}(x_3-x_5).
\end{equation}
and integrand in \eqref{eq:intboxing} vanishes. When $d>4$ in general, such a delta function vanishes and the contribution transfers to the integrand in \eqref{eq:intboxing}. Thus the lower-loop contribution behaves as the $1/\epsilon$ divergence in integrals with double pole in dimensional regularization.

Such a calculation directly applies for external vertices which appear in more than one propagators. For example, in 4d conformal integrals, the next simplest case is
\begin{equation}\label{eq:boxing2}
    \square_{3}\frac{x_{37}^2}{x_{35}^2x_{36}^2}=-4\pi^2\frac{x_{37}^2}{x_{36}^2}\delta^{(4)}(x_3-x_5)-4\pi^2\frac{x_{37}^2}{x_{35}^2}\delta^{(4)}(x_3-x_6)+4\frac{x_{36}^2x_{57}^2+x_{35}^2x_{67}^2-x_{37}^2x_{56}^2}{(x_{35}^2)^2(x_{36}^2)^2}.
\end{equation}
This can be generalized to arbitrary case. If we concentrate on the lower-loop contributions for singularity information in deeper depth of svMPLs, then the boxing can be readily turned into a graph reduction rule by formula as in \eqref{eq:boxing1} and \eqref{eq:boxing2}.

\section{Empirical rule for the ansatz of conformal integrals}\label{app:reducerule}
For rigid conformal integrals, we have observed in several examples that if there are new letters other than 0 and 1 in the results, they will also appear as the leading singularities of those reduced graphs of original conformal integral, and they only appear in the last two entries of the svMPLs\footnote{We note that the definition of svMPLs involving general $a_n$ beyond 0 and 1 may differ in literature. We stick to the definition in~\cite{Schnetz:2013hqa,Schnetz:2021ebf}.}. This provides us extra information about the function space involved. Before further stating the reduction rule in detail, we must emphasize that this is only an observation rather than a principle for the construction of ansatz. It is justified by successfully bootstrapping the final results.

\begin{figure}[htbp]
    \centering
    \includegraphics[width=0.5\textwidth]{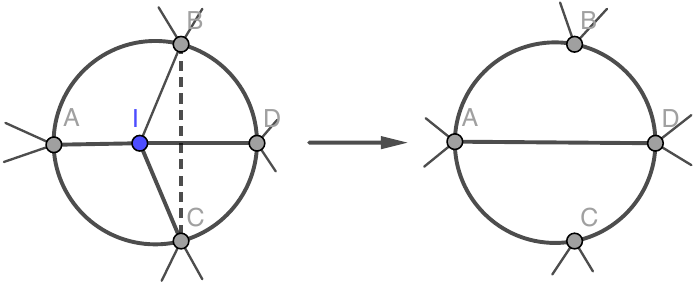}
    \caption{The reduction of conformal integrals by multiplying an inverse box. `I' denotes an internal (loop) vertex and remaining vertices can either be external or internal. It is a standard boxing if A,B,C or B,C,D are all external points and A or D does not connect to other points.}\label{fig:gboxing}
\end{figure}

The reduced graphs are derived from the original conformal integral as indicated in Fig.~\ref{fig:gboxing}.
Let us take the three-loop hard and easy integrals as examples. They are both rigid and involve a common leading singularity $u/(1-u)/(z-\bar{z})$. They also have another leading singularity $1/(z-\bar{z})$ and $u/(z-\bar{z})^2$ respectively. We can not deduce the ansatz for each of them from their own leading singularities. Now we include the leading singularities analysis of the reduced graphs for them. On the one hand, the reduced graph for easy integral only involve $1/(z-\bar{z})$, which is the standard leading singularity for conformal integrals. Meanwhile, the reduced graphs for hard integral only involves leading singularity $1/(z-\bar{z})^2$, which acquires another $1/(z-\bar{z})$ upon the standard leading singularity. On the other hand, the function space for easy integral only involves svHPLs and the function space for hard integral involves svMPLs with new letter $\bar{z}$ appearing in the last two entries. This alignment between leading singularity of reduced graph and the corresponding function space for original integral has an amazing correspondence with the boxing differential equation\footnote{For an $L$-loop conformal integral $\Phi^{(L)}\equiv f^{(L)}/(z-\bar{z})$ satisfying boxing differential equation, we will have $z\bar{z}\partial_{z}\partial_{\bar{z}}f^{(L)}=cf^{(L-1)}$ or $(1-z)(1-\bar{z})\partial_{z}\partial_{\bar{z}}f^{(L)}=cf^{(L-1)}$ depending on the normalization, where $c$ is some constant.}. Since the boxing differential equation reduces an $l$-loop conformal integral to $l{-}1$-loop one and when we try to solve the boxing differential equation, $1/(z-\bar{z})^2$ in the lower-loop conformal integral will naturally give us a new letter $z-\bar{z}$ and they appear only in the last two entries of svMPLs. In a similar way, if there is a leading singularity $1/(1-z\bar{z})/(z-\bar{z})$ in the reduced lower-loop integral, then solving the boxing differential equation will give us a new letter $z-1/\bar{z}$. As a reminder, the reduction rule we introduced in Fig.~\ref{fig:gboxing} is not boxing in general. Therefore, such a correspondence is nontrivial.

An important example of this empirical rule is the bootstrapping of $\mathcal{I}^{(a,b,c)}_{176}$ in \eqref{eq:I4L176abc}. Since $\mathcal{I}^{a}_{176}$ does not have a simple graphical representation, it is composite. We do not apply reduction rule to it directly, but apply the reduction rule to its components, $\mathcal{I}^{(4)}_{176}, \mathcal{I}^{b}_{176}, \mathcal{I}^{c}_{176}$, separately\footnote{Though we decompose $\mathcal{I}^{(4)}_{176}$ into three part $\mathcal{I}^{a,b,c}_{176}$, but from the view of integrand, $\mathcal{I}^{a}_{176}$ is composed of $\mathcal{I}^{(4)}_{176}$ and $\mathcal{I}^{b,c}_{176}$. The former is organized by leading singularity, that is, the integral results. The latter is organized by integrands.}. Using this rule, we conjecture that $\mathcal{I}^{b}_{176},\mathcal{I}^{(4)}_{176}$ should have $1/\bar{z}$ in the last two entries of svMPLs, while $\mathcal{I}^{c}_{176}$ does not have this new letter, that is, it is totally composed of svHPLs. The final bootstrap results indeed verify this conjecture.

Although we can use such a correspondence proposed above to predict the possible function space by the leading singularities of reduced graphs, we have to point out that their usage is very limited. The main reason is that it applies to rigid conformal integrals. We find that in four loops, most conformal integrals with graph representations are either not rigid or not free of elliptic cut. In many cases, such a conformal integral contains many different cuts. On the support of some cut, it may appear to be rigid, but on another cut it contains a double pole or appears to be an elliptic integral. This is the reason why we have not accumulated a lot of data to further test this observation. Secondly, since this reduction requires such a special structure of the integrand, it certainly does not apply to all conformal integrals. Nevertheless, it works fine for some examples we are studying.

\section{Three loop conformal integrals basis}\label{app:threeloopint}
In this appendix, we draw the 15 three-loop integral basis generated from $f$-graphs including planar and nonplanar ones. They form a basis in the integrand level. However, there are also additional integral identities between them. See Sec.~\ref{sec:threeloopconformal} for more details. At last, there are 12 independent functions remaining. We also provide the data of integrands in ancillary files.
\begin{figure}[htbp]
    \centering
    \renewcommand{\arraystretch}{1.05}
    \begin{tabular}{ccccc}
        \raisebox{-0.5\height}{\includegraphics[scale=0.28]{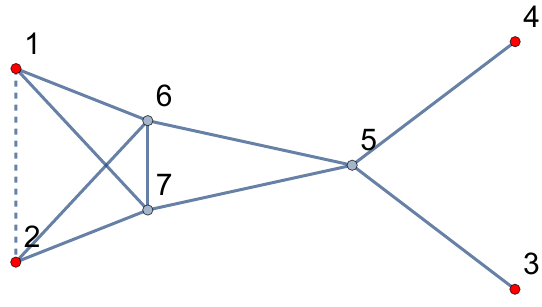}} &
        \raisebox{-0.5\height}{\includegraphics[scale=0.28]{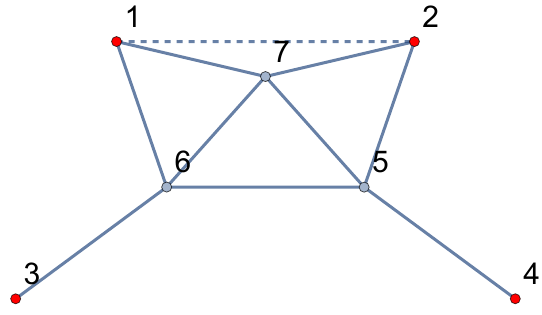}} &
        \raisebox{-0.5\height}{\includegraphics[scale=0.28]{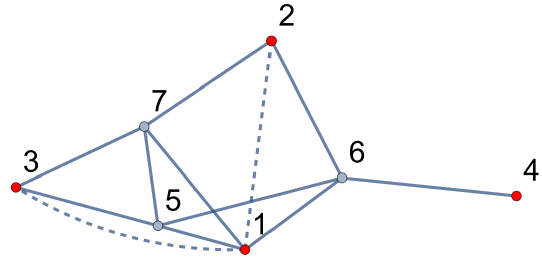}} &
        \raisebox{-0.5\height}{\includegraphics[scale=0.28]{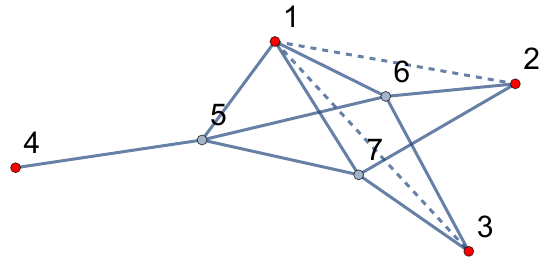}} &
        \raisebox{-0.5\height}{\includegraphics[scale=0.28]{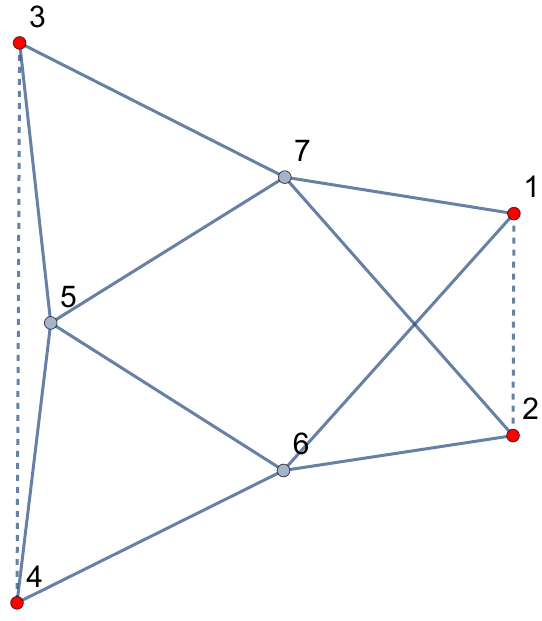}} \\[-2pt]
        \small $I^{(3)}_1$ & \small $I^{(3)}_2$ & \small $I^{(3)}_3$ & \small $I^{(3)}_4$ & \small $I^{(3)}_5$ \\[6pt]
        \raisebox{-0.5\height}{\includegraphics[scale=0.28]{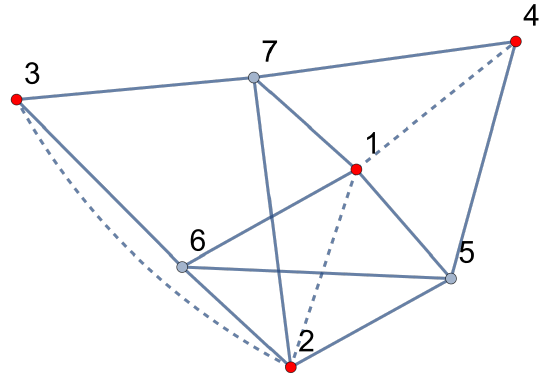}} &
        \raisebox{-0.5\height}{\includegraphics[scale=0.28]{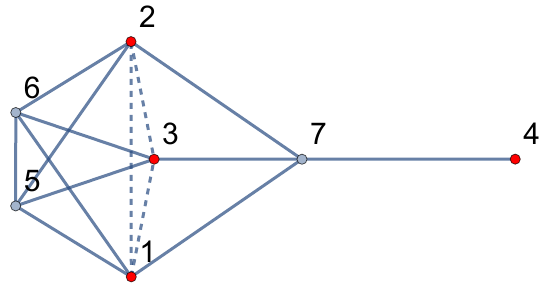}} &
        \raisebox{-0.5\height}{\includegraphics[scale=0.28]{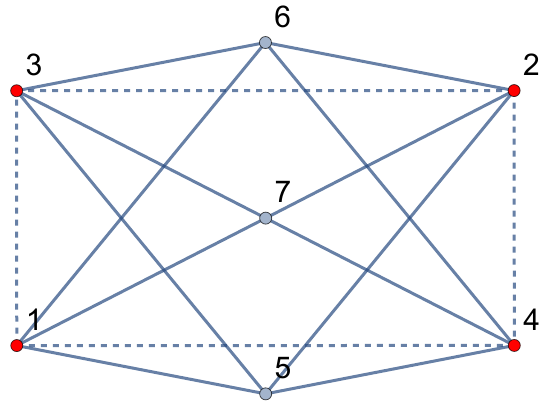}} &
        \raisebox{-0.5\height}{\includegraphics[scale=0.28]{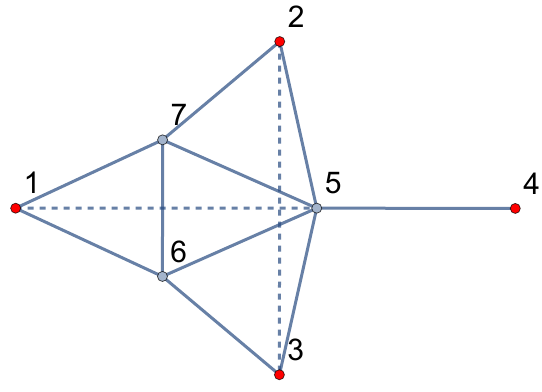}} &
        \raisebox{-0.5\height}{\includegraphics[scale=0.28]{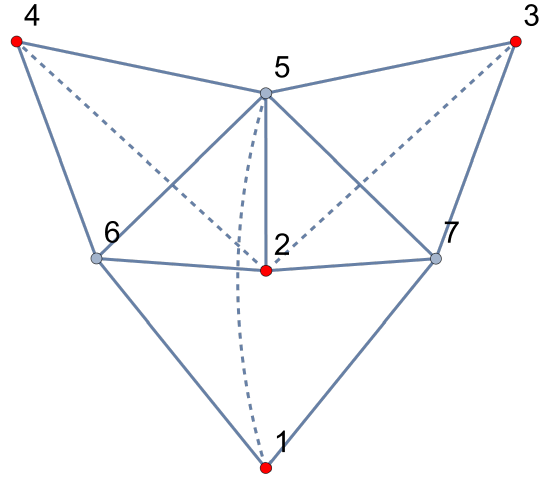}} \\[-2pt]
        \small $I^{(3)}_6$ & \small $I^{(3)}_7$ & \small $I^{(3)}_8$ & \small $I^{(3)}_9$ & \small $I^{(3)}_{10}$ \\[6pt]
        \raisebox{-0.5\height}{\includegraphics[scale=0.28]{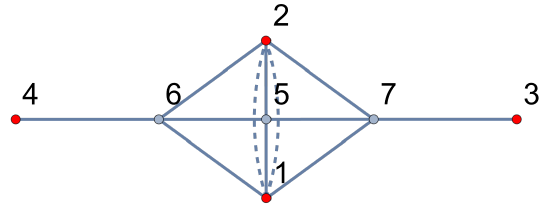}} &
        \raisebox{-0.5\height}{\includegraphics[scale=0.28]{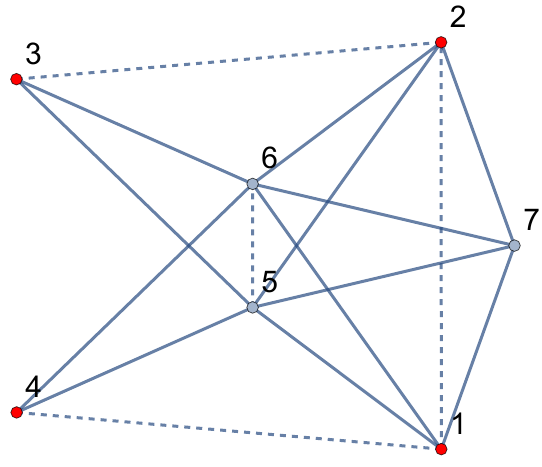}} &
        \raisebox{-0.5\height}{\includegraphics[scale=0.28]{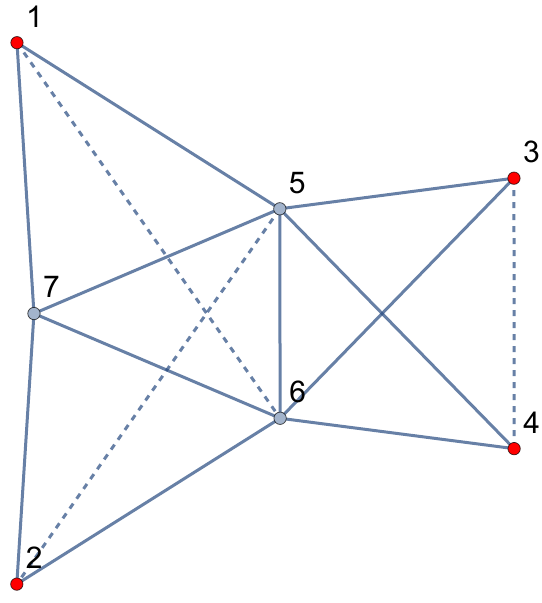}} &
        \raisebox{-0.5\height}{\includegraphics[scale=0.28]{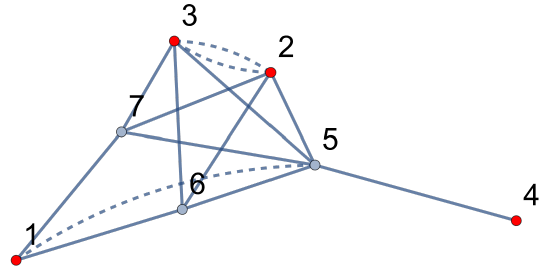}} &
        \raisebox{-0.5\height}{\includegraphics[scale=0.28]{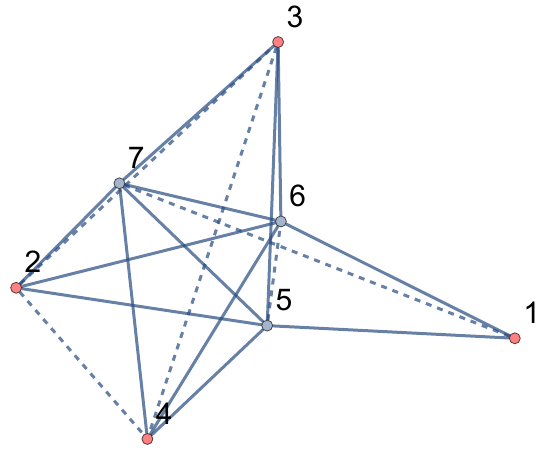}} \\[-2pt]
        \small $I^{(3)}_{11}$ & \small $I^{(3)}_{12}$ & \small $I^{(3)}_{13}$ & \small $I^{(3)}_{14}$ & \small $I^{(3)}_{15}$
    \end{tabular}
    \caption{The 15 three-loop conformal integral basis generated from $f$-graphs. $\mathcal{I}^{(3)}_{2}$ is related to $\mathcal{I}^{(3)}_{3}$ and $\mathcal{I}^{(3)}_{9}$ is related to $\mathcal{I}^{(3)}_{11}$ by magic identities. $\mathcal{I}^{(3)}_{15}$ is related to the rest of the integrals by Gram identity. $\mathcal{I}^{(3)}_{6,7,8}$ are three reducible integrals that are products of lower-loop ones.}
    \label{fig:threeloopbasis}
\end{figure}




\bibliographystyle{JHEP}
\bibliography{biblio}

\end{document}